\documentclass[journal]{IEEEtran}
\usepackage{color}
\usepackage{amsmath}
\usepackage{amssymb}
\usepackage{amsthm}
\usepackage{amsfonts}
\usepackage{multirow}
\usepackage[table,xcdraw]{xcolor}
\usepackage{hyperref}

\ifCLASSINFOpdf
\usepackage[pdftex]{graphicx}
\else
\fi
\usepackage[ruled,norelsize]{algorithm2e}

\makeatletter
\newcommand{\removelatexerror}{\let\@latex@error\@gobble}
\makeatother

\usepackage{subfigure}
\DeclareMathOperator*{\argmax}{argmax}

\begin{document}
%
\title{Multiple Sound Source Localisation with Steered Response Power Density and Hierarchical Grid Refinement}
%
%
%
\author{Mert~Burkay~\c{C}\"{o}teli,~\IEEEmembership{Student Member,~IEEE}, Orhun~Olgun,~\IEEEmembership{Student Member,~IEEE}, and~H\"{u}seyin~Hac\i{}habibo\u{g}lu,~\IEEEmembership{Senior Member,~IEEE}
\thanks{M.~B.~\c{C}\"{o}teli, O.~Olgun, and H.~Hac\i{}habibo\u{g}lu are with the  Spatial Audio Research Group (SPARG), Graduate School of Informatics, Middle East Technical University (METU), Ankara, TR-06800, Turkey e-mail: \{mbcoteli, oolgun,  hhuseyin\}@metu.edu.tr.}
\thanks{The work reported in this paper is supported by the Turkish Scientific and Technological Research Council (T\"{U}B\.{I}TAK) Research Grant 113E513 ``Spatial Audio Reproduction Using Analysis-based Synthesis Methods''.}
\thanks{Manuscript received March 03, 2018.}}

%
%

\markboth{SUBMITTED FOR PUBLICATION IN IEEE/ACM Transactions on Audio, Speech and Language Processing,~Vol.~XX, No.~YY, March~2018}%
{\c{C}\"{o}teli, Olgun, and Hac\i{}habibo\u{g}lu: Multiple Source Localization with Rigid Spherical Microphone Arrays}
%



\maketitle

\begin{abstract}
Estimation of the direction-of-arrival (DOA) of sound sources is an important step in sound field analysis. Rigid spherical microphone arrays allow the calculation of a compact spherical harmonic representation of the sound field. A basic method for analysing sound fields recorded using such arrays is steered response power (SRP) maps wherein the source DOA can be estimated as the steering direction that maximises the output power of a maximally-directive beam. This approach is computationally costly since it requires steering the beam in all possible directions. This paper presents an extension to SRP called steered response power density (SRPD) and an associated, signal-adaptive search method called hierarchical grid refinement (HiGRID) for reducing the number of steering directions needed for DOA estimation. The proposed method can localise coherent as well as incoherent sources while jointly providing the number of prominent sources in the scene. It is shown to be robust to reverberation and additive white noise. An evaluation of the proposed method using simulations and real recordings under highly reverberant conditions as well as a comparison with state-of-the-art methods are presented. 
\end{abstract}


%
\IEEEpeerreviewmaketitle

\section{Introduction}\label{sec:intro}

Sound source localization is an essential stage in sound field analysis and is used in a multitude of contexts including object-based audio, robot audition, and acoustic surveillance. Spatial information encapsulated in the sound field provides a basis to localize sound sources. Such information can be extracted from recordings made with  microphone arrays~\cite{brandstein2013microphone}.

Recent years have seen progress in the development of the theory and applications of rigid spherical microphone arrays~\cite{Rafaely:2015do}. Different methods developed for DOA estimation using such arrays include those which are specific to array signal processing, subspace-based methods adapted from classical spectrum estimation theory, and methods that are based on the energetic analysis of sound fields~\cite{Jarrett:2016ie}. These methods suffer from at least one of the following problems: They either have a high computational cost, require prior information about the number of sources, or fail when there are multiple coherent sources or a high level of reverberation. 

Steered-response power (SRP) method\footnote{It should be noted that SRP method that we are referring to in this article, is based on plane-wave decomposition (PWD) using a maximally-directive beampattern and not for example minimum-variance distortionless response beamforming the eigenbeam domain (EB-MVDR)~\cite{sun2011robust}.} is based on the calculation of a spatial map of sound power calculated at different steering directions. Source DOAs correspond to the maxima of this distribution and can be obtained after the SRP map is calculated. Although SRP maps can provide excellent DOA estimation accuracy, two important problems prevent their widespread use in DOA estimation: 1) computational cost~\cite{Jarrett:2010vs}, and 2) robustness to additive noise~\cite{sun2011robust}.

A new DOA estimation and source counting method robust to reverberation and additive noise that aims to reduce the computational cost of steered beamforming based sound source localization is proposed in this article. First, a measure of the spatially averaged steered response power, called the steered response power density (SRPD) is defined. SRPD is then interpreted as the probability of a sound source being present in a spherical quadrilateral sector and used in the adaptive refinement of a hierarchical search grid. This is done by selecting the new steering directions based on the change in the total information gain. Only the directional sectors that are likely to contain a source are selected and scanned, allowing DOA estimation without calculating the SRP map on the whole sphere. This is followed by the segmentation and labelling of the multiresolution SRPD maps, allowing the joint localization and counting of the prominent sources in the acoustic scene. The effectiveness of the algorithm is demonstrated via an extensive localization experiment. Operation of the proposed method under real-life conditions is demonstrated. A comparison of the proposed method with three state-of-the-art methods as well as the baseline SRP method is also presented.
 
The paper is structured as follows. A summary of earlier research using spherical microphone arrays for DOA estimation is given in Sec.~\ref{sec:intro}. A brief introduction to the theory of rigid spherical microphone arrays is given in Sec.~\ref{sec:rma}. Steered response power density (SRPD) is defined and its calculation from microphone array recordings is described in Sec.~\ref{sec:spm}. A method using a signal-adaptive hierarchical search based on spatial entropy is proposed in Sec.~\ref{sec:method}. The computational cost of the proposed method is compared with that of SRP in Sec.~\ref{sec:comp}. The evaluation of the proposed method under different emulated conditions is given, a comparison with the state-of-the-art methods is presented, and the utility of the method under real-life conditions is demonstrated in Sec.~\ref{sec:eval}. Sec.~\ref{sec:conc} concludes the paper. 

\section{Background}\label{sec:intro}

One of the most straightforward ways of DOA estimation using rigid spherical microphone arrays involves scanning all possible directions by steering a maximum directivity beam and calculating a map of the response power for each of these directions~\cite{sun2012localization}. The maxima of this map occur at the DOAs of the sound sources present in the recorded scene. This brute-force approach, called steered response power (SRP) map, incurs a computational cost that can be prohibitively high for practical applications. 

The use of SRP for sound source localisation is also common for general microphone arrays which may or may not have a regular geometric structure (e.g. linear or planar arrays). In such cases, SRP maps are typically obtained via generalised cross correlation with phase transform (GCC-PHAT)~\cite{knapp1976generalized} or its derivatives. Several different methods that aim to reduce the computational cost associated with this approach have been proposed. For example, stochastic region contraction (SRC)~\cite{Do:2007ut} and coarse-to-fine region contraction (CFRC)~\cite{Do:2007ka} are two methods that search for the volumetric region with the global SRP maximum by gradually shrinking the search grid boundaries. Other methods search for the global SRP maximum from an SRP map obtained via a spatially averaged GCC-PHAT by iteratively decomposing the search volume into subvolumes~\cite{Marti:2013jm, Lima:2015kr} or use bound and branch method to subdivide the search grid to obtain the global SRP maximum~\cite{Nunes:2014ii}.

Subspace-based DOA estimation methods such as multiple signal classification in the eigenbeam domain (e.g. EB-MUSIC) and estimation of signal parameters via rotational invariance in the eigenbeam domain (EB-ESPRIT)~\cite{teutsch2008detection, sun2011robust} are adaptations of the well-known spectrum estimation methods to the problem of DOA estimation. These methods provide accurate DOA estimations. However, they require \textit{a priori} information on the number of sources and are computationally costly. Also, while EB-ESPRIT has a lower computational cost in comparison with EB-MUSIC, it may still fail for directions where the estimation function is singular~\cite{sun2012localization}. 

Methods that aim to reduce the computational cost of DOA estimation with rigid spherical microphone arrays have also been proposed. One of these methods, inspired by energetic analysis of sound fields, uses pseudo-intensity vectors (PIV) calculated from zeroth and first-order spherical harmonics~\cite{Jarrett:2010vs, Evers:2014fq}. While the PIV method was shown to be successful in localizing multiple sources when these sources are approximately W-disjoint orthogonal~\cite{rickard2002approximate} (e.g. speech), its performance will decrease for coherent sources. Such a situation may occur, for example, in cases with multiple musical instruments or a source and its strong specular reflections. More recently, two extensions to the PIV method have been proposed in order to improve its estimation accuracy. Augmented intensity vectors (AIV) method extends the PIV concept using higher-order spherical harmonics~\cite{hafezi2016multiple,hafezi2017augmented}. Subspace PIV (SSPIV) uses DOA estimations obtained from PIVs in the signal subspace calculated from the time-frequency smoothed spatial covariance matrix via singular value decomposition (SVD)~\cite{moore2017direction}. Similarly, a multisource estimation consistency metric, which provides a measure of estimation confidence, can be used to improve accuracy for multiple source scenarios~\cite{hafezi2017multi}.

The desire to exploit the DOA estimation accuracy of EB-MUSIC prompted the investigation of ways to decrease the computational cost it incurs. One possibility to achieve this aim is to estimate DOAs using EB-MUSIC only at time-frequency bins for which only a single source is active~\cite{khaykin2009coherent, Khaykin:2012fd}. For such bins, it is possible to deduce the dimensions of the signal and noise subspaces and also to reduce the number of time-frequency bins over which EB-MUSIC spectrum is to be calculated. Direct-path dominance (DPD) test uses the ratio of the largest two singular values of the time-frequency smoothed spatial covariance matrix in order to identify bins with only one active source~\cite{nadiri2014localization}. DPD-test was also used as a pre-processing stage for the PIV method~\cite{moore2015direction}.

While DPD test identifies time-frequency bins that contain a single source, it is also possible to use the correlation coefficient between microphone pairs to identify directional zones that contain a single source and search for sources (e.g. by steered beamforming) in those zones only~\cite{6557035, 7471644}. This approach is conceptually similar to the method proposed in this article. 

\section{RIGID SPHERICAL MICROPHONE ARRAYS}\label{sec:rma}
A function defined on the unit sphere can be represented using the spherical harmonic functions as a basis such that:
\begin{equation}
f(\theta,\phi)=\sum_{n=0}^{\infty}\sum_{m=-n}^{n}f_{nm}Y_n^m(\theta,\phi).
\end{equation}
Here, the spherical harmonic functions of order $n\in\mathbb{N}$ and degree $m\in\mathbb{Z}$ are defined as:
\begin{equation}
Y_n^m(\theta,\phi)=\sqrt{\frac{2n+1}{4\pi}\frac{(n-m)!}{(n+m)!}}P_n^m(\cos{\theta})e^{\textrm{i}m\phi}
\end{equation}
with $n\geq{}0$, $m\leq|n|$, where $P_n^m(\cdot)$ is the associated Legendre functions, $\theta$ and $\phi$ are the inclination and azimuth angles, and
\begin{equation}
f_{nm} = \int_{0}^{2\pi}\int_{0}^{\pi}f(\theta,\phi)[Y_n^m(\theta,\phi)]^*\sin\theta d\theta d\phi\label{eq:SHD}
\end{equation}
are the corresponding spherical harmonic coefficients, respectively. 

The projection of the function, $f(\theta,\phi)$, onto spherical harmonic basis as described above is called the spherical harmonic decomposition (SHD). Since SHD results in a compact representation of bandlimited functions and distributions on a sphere using only a few non-zero SHD coefficients, it is used in many different fields of physics as well as in signal processing including applications in acoustics~\cite{Williams:1999wx}. 

If the pressure distribution is known at each point on a sphere, (\ref{eq:SHD}) can be used to obtain the SHD directly. However, in real-life applications pressure can be sampled only at a finite number of points on the sphere. If the pressure is sampled at $Q\geq{}(N+1)^2$ discrete points, $(\theta_q,\phi_q)$, the spherical harmonic components can be calculated using the corresponding spherical quadrature up to a maximum degree of $n=N$ as:
\begin{equation}
p_{nm}(k) = \sum_{q=1}^{Q}w_q p(\theta_q,\phi_q,k)[Y_n^m(\theta_q,\phi_q)]^*\label{eq:DSHD}
\end{equation}
where $k=2\pi{}f/c$ is the wave number, $c$ is the speed of sound, and $w_q$ are the quadrature weights.  In order for this expression to converge to the real SHD coefficients, spherical sampling scheme has to satisfy the discrete orthonormality condition~\cite{li2004flexible}.


A spatially bandlimited approximation of the pressure around a rigid sphere of radius $r_a$, located at the origin, due to a unit amplitude plane wave incident from the direction $(\theta_S, \phi_S)$ and with a frequency $f$, can be written for $r\geq{}r_a$ as:

\begin{equation}
p(\theta,\phi,k)=\sum_{n=0}^{N}\sum_{m=-n}^{n}p_{nm}(k)Y_n^m(\theta,\phi),
\end{equation}
where 
\begin{equation}
p_{nm}(k) =4\pi{}i^nb_n(kr_a)[Y_n^m(\theta_S,\phi_S)]^*
\end{equation}
with 
\begin{equation}
b_n(kr)=j_n(kr)-\dfrac{j_n^{'}(kr_a)}{h_n^{(2)'}(kr_a)}h_n^{(2)}(kr).
\end{equation}
Here, $j_n(\cdot)$, $h_n^{(2)}(\cdot)$, $j_n^{'}(\cdot)$, and $h_n^{(2)'}(\cdot)$ are, the spherical Bessel function of the first kind, spherical Hankel function of the second kind and their derivatives with respect to their arguments, respectively~\cite{Rafaely:2015do}. It may be observed that an important benefit of this representation is the decoupling of direction and frequency dependent terms. 

Rigid spherical microphone arrays consist of a number of pressure sensitive microphones positioned at appropriate quadrature nodes on the surface of a rigid spherical baffle. Such arrays have gained popularity in acoustic scene analysis applications as they possess spherical symmetry and allow for an almost trivial calculation of spherical harmonic coefficients subject to the order limitations mentioned above. Spatial aliasing that results from using a finite number of samples on the sphere is discussed elsewhere~\cite{rafaely2007spatial}.

\section{STEERED RESPONSE POWER DENSITY}\label{sec:spm}
In a realistic scenario where multiple sources or a source and its reflections are present, the SHD coefficients will consist of a linear combination of multiple directional terms:
\begin{equation}
p_{nm}(k) = 4\pi{}i^nb_n(kr_a)\sum_{s=1}^{S}\alpha_s(k)[Y_n^m(\theta_s,\phi_s)]^*
\end{equation}
where $\alpha_s(k)\in\mathbb{C}$ is the amplitude of a single plane wave. Using the completeness property of the spherical harmonic functions~\cite{Rafaely:2015do}, it can be shown that:
\begin{eqnarray}
y(\theta,\phi,k)&=&\sum_{n=0}^{\infty}\sum_{m=-n}^{n}\frac{p_{nm}(k)}{4\pi{}i^nb_n(kr_a)}Y_n^m(\theta,\phi)\\&=&\sum_{s=1}^S\alpha_s\delta(\cos\theta-\cos\theta_s)\delta(\phi-\phi_s)\nonumber,
\end{eqnarray}
where $\delta(\cdot)$ is the Dirac delta function.

In other words, if there is no order limitation on the SHD, the above operation will result in a combination of a finite number, $S$ of Dirac impulses on the unit sphere with complex amplitudes, $\alpha_s(k)$, located at directions that correspond to the directions of the individual plane waves, $(\theta_s, \phi_s)$. This is called \textit{plane wave decomposition} (PWD)~\cite{Rafaely:2004ez}. 

Only a spatially bandlimited approximation of PWD can be obtained with  rigid spherical microphone arrays due to the order limitation mentioned above, such that: 
\begin{equation}
y_N(\theta,\phi,k)=\sum_{n=0}^{N}\sum_{m=-n}^{n}\frac{p_{nm}(k)}{4\pi{}i^nb_n(kr_a)}Y_n^m(\theta,\phi)\label{eq:yN}
\end{equation}
It may then be shown by the spherical harmonics addition theorem~\cite{Rafaely:2010ih} that the effect of order limitation results in spatially bandlimited impulses:
\begin{equation}
y_N(\theta,\phi,k)=\frac{N+1}{4\pi}\sum_{s=1}^S\alpha_s(k)\left[\frac{P_{N+1}(\cos\Theta_s)-P_{N}(\cos\Theta_s)}{P_{1}(\cos\Theta_s)-P_{0}(\cos\Theta_s)}\right]\nonumber\label{eq:SRP}
\end{equation}
where $\Theta_s$ is the angle between the direction of the incident plane wave, $(\theta_s, \phi_s)$ and the analysis direction, $(\theta, \phi)$, and $P_n(\cdot)$ is the Legendre polynomial of order $n$. The resulting beam pattern is called the \textit{regular beam pattern}~\cite{li2007flexible} or \textit{plane-wave decomposition beam pattern}~\cite{Rafaely:2010ih} and is maximally directive.

The functional, $y_N(\theta,\phi,k)$ is called the \textit{steered response power} (SRP). When it is interpreted as a distribution on the unit sphere, the spatial detail that it can resolve, also known as the \textit{Rayleigh condition}, is approximately $\pi/N$~\cite{Rafaely:2004ez}. In other words, only sources that have more than $\pi/N$ separation between their DOAs can be discriminated as separate, independent sources.

DOA estimation using SRP involves finding the steering direction, $(\theta_u, \phi_u)$ that maximises the power, such that:
\begin{equation}
(\theta_u, \phi_u)_k=\argmax_{\theta_u,\phi_u}|y_N(\theta_u,\phi_u,k)|^2.
\end{equation}

For multiple sources, the SRP functional will have multiple peaks corresponding to source DOAs. These can be accurately identified by using a dense search grid with a desired resolution. However, this brute-force approach is usually not suitable for practical applications due to its high computational cost. 

For the case where a sparse, multiresolution search grid is to be used, SRP can miss peaks especially at low resolutions. A more appropriate functional which can be used for source localization is proposed in this article. We define \textit{steered response power density} (SRPD) for an arbitrary, bounded surface element, $\mathcal{S}_i$, on the unit sphere as:
\begin{equation}
	\mathcal{P}_i(k)=\frac{1}{A_i}\int_{\mathcal{S}_i}|y_N(\theta,\phi,k)|^2d\mathcal{S}_i
\end{equation} 
where $A_i\triangleq{}A(\mathcal{S}_i)$ is the area of the surface element. Note that SRP is related to SRPD as:
\begin{equation}
	\lim_{A_i\rightarrow{}0}\mathcal{P}_i(k)=|y_N(\theta_i,\phi_i,k)|^2
\end{equation}
where $(\theta_i, \phi_i)$ is the centre of the differential region, $\mathcal{S}_i$. 

It is possible to express SRPD using (\ref{eq:yN}) as:
\begin{equation}
	\mathcal{P}_i(k)=\sum_{n,m,n',m'}\frac{p_{nm}(k)p_{n'm'}^*(k)}{b_n(kr_a)b_{n'}^*(kr_a)}Q_{n,n'}^{m,m'}(\mathcal{S}_i)\label{eq:SRPD}
\end{equation}
where:
\begin{equation}
Q_{n,m}^{n',m'}(\mathcal{S}_i)=\frac{1}{(4\pi)^2A_i}\int_{\mathcal{S}_i}Y_n^m(\theta,\phi)\left[Y_{n'}^{m'}(\theta,\phi)\right]^*d\mathcal{S}_i.	
\end{equation}

Here, the decoupling of the frequency and the direction-dependent terms via SHD allows isolating the integration to terms which do not depend on the SHD coefficients but only on the analysis region.

It is possible to express the sum in (\ref{eq:SRPD}) as the grand sum of the matrix: 
\begin{equation}
\mathbf{H}_i=\mathbf{P}\circ{}\mathbf{Q}_i	
\end{equation}
 where $\circ$ represents the Hadamard (i.e.~element-wise) product, $\mathbf{P}=\mathbf{p}\mathbf{p}^H$ is an $(N+1)^2\times{}(N+1)^2$ matrix with 
 
\begin{equation}
\mathbf{p}=\left[\frac{p_{00}(k)}{4\pi{}b_0(kr_a)},\ \frac{p_{1-1}(k)}{4\pi{}b_1(kr_a)}, \cdots{},\frac{p_{NN}(k)}{4\pi{}b_N(kr_a)}\right]^T
\end{equation}
and the  $(N+1)^2\times{}(N+1)^2$ cross spatial density matrix $\mathbf{Q}_i$ is given as:

\begin{equation}
\mathbf{Q}_i=
\begin{bmatrix}
	Q_{0,0}^{0,0}(\mathcal{S}_i) & Q_{0,0}^{1,-1}(\mathcal{S}_i) &  \cdots & Q_{0,0}^{N,N}(\mathcal{S}_i) \\
	Q_{1,-1}^{0,0}(\mathcal{S}_i) & Q_{1,-1}^{1,-1}(\mathcal{S}_i) &  \cdots & Q_{1,-1}^{N,N}(\mathcal{S}_i)\\
	\vdots & \vdots  & \ddots & \vdots\\
	Q_{N,N}^{0,0}(\mathcal{S}_i) & Q_{N,N}^{1,-1}(\mathcal{S}_i) &  \cdots & Q_{N,N}^{N,N}(\mathcal{S}_i)
\end{bmatrix}.
\end{equation}
Note that both $\mathbf{P}$ and $\mathbf{Q}_i$ are Hermitian. 

The grand sum of $\mathbf{H}_i$ can be represented as:
\begin{equation}
	\mathcal{P}_i=\mathbf{e}^T\mathbf{S}_i\mathbf{e}=\mathbf{e}^T\left(\mathbf{P}\circ{}\mathbf{Q}_i	\right)\mathbf{e}
\end{equation}
where $\mathbf{e}$ is a column vector of ones. This expression can be simplified by employing an identity of the Hadamard product~\cite{Horn:1990tf} and the eigendecomposition of the cross spatial density matrix such that:
\begin{equation}
	\mathcal{P}_i=\mathrm{tr}(\mathbf{P}\mathbf{Q}_i^T)=\mathrm{tr}(\mathbf{P}\mathbf{V}_i\mathbf{D}\mathbf{V}_i^H)
\end{equation}
where $\mathrm{tr}(\cdot)$ is the trace operator. Here, the columns of $\mathbf{V}_i$ are the eigenvectors, and the diagonal matrix $\mathbf{D}$ contains the eigenvalues $\lambda_{i,m}$ of $\mathbf{Q}_i^T$. Using the cyclic permutation invariance property of the trace operator~\cite{Arfken:2005th}:
\begin{eqnarray}
	\mathcal{P}_i&=&\mathrm{tr}(\mathbf{V}_i^H\mathbf{P}\mathbf{V}_i\mathbf{D})\\&=&\mathbf{e}^T\left(\mathbf{V}_i^H\mathbf{P}\mathbf{V}_i\circ\mathbf{D}_i	\right)\mathbf{e}
\end{eqnarray}

Since the eigenvalue matrix $\mathbf{D}_i$ is diagonal, the grand sum of the Hadamard product will only contain a sum of the diagonal elements of $\mathbf{V}_i^H\mathbf{P}\mathbf{V}_i$ weighted by the eigenvalues, $\lambda_{i,m}$ of the cross spatial density matrix, such that the SRPD is given as:
\begin{equation}
	\mathcal{P}_i=\sum_m{}\lambda_{i,m}\left[\mathbf{V}_i^H\mathbf{P}\mathbf{V}_i\right]_{m,m},\label{eq:srpd}
\end{equation}
or more compactly as:
\begin{equation}
	\mathcal{P}_i=\|\mathbf{V}_i^H\mathbf{p}\|_{\mathbf{D}_i}^2\label{eq:srpdwm}
\end{equation}
where $\|\mathbf{x}\|_\mathbf{D}=(\mathbf{x}^H\mathbf{D}\mathbf{x})^{1/2}$ represents the weighted norm. In other words, once the SHD coefficients are known, SRPD can be calculated using the eigenvalues and the eigenvectors of the cross spatial density matrix, $\mathbf{Q}_i$.

The expression above assumes that all the eigenvalues and eigenvectors of the spatial density matrix are employed. However, most of the eigenvalues will be very small depending on the area over which the SRPD is calculated. Fewer eigenvalues and eigenvectors can be used in order to reduce the computational cost associated with the calculation of SRPD. Defining an energy ratio as:
\begin{equation}
	L_{i,M}\triangleq\sum_{m=0}^{M}\lambda_{i,m}^2\bigg/\sum_{m=0}^{(N+1)^2}\lambda_{i,m}^2
\end{equation}
the largest $M<(N+1)^2$ eigenvalues and the corresponding eigenvectors can be selected, for which this ratio is greater than a given threshold close to unity. This way the computational cost can be substantially reduced without having a detrimental effect on the calculated SRPD for the corresponding area. In the examples that we present in the article, we use the largest $M$ eigenvalues and eigenvectors such that $L_{i,M}\geq{}0.99$.

The most important difference between SRP and SRPD is that of coverage: while SRP corresponds to the response power for a distinct steering direction,  SRPD corresponds to the power density in an area of interest. As will be evident in the next section, the latter is more suitable when a multiresolution search grid is used for DOA estimation. 

\section{SPATIAL ENTROPY BASED GRID REFINEMENT}\label{sec:method}
\begin{figure*}[!t]
\centering
\includegraphics[width=\textwidth]{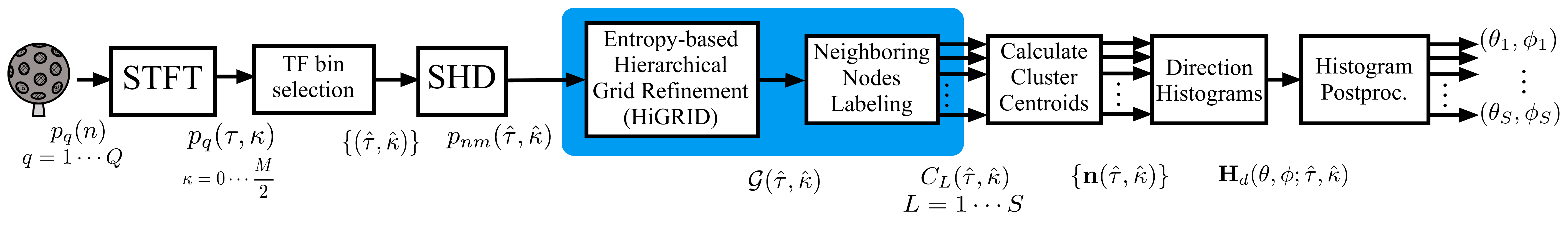}
\caption{The flow diagram of the proposed algorithm. The core part is highlighted with a blue box.}\label{fig:flow}
\end{figure*}

The method proposed in this paper aims to provide accurate DOA estimations comparable with those provided by SRP but with a lower computational cost, for multiple, possibly coherent sources. This is attained by using a hierarchical, multiresolution search grid which is refined based on information gain. 

The proposed method is based on the clustering of local DOA estimates obtained from individual bins in the time-frequency domain. For this purpose, short-time Fourier transforms (STFT) of each microphone channel from the spherical microphone array are obtained first. This is followed by a time-frequency bin selection process that chooses bins which are likely to contain one or more strong direct path components. We use time-frequency bins that correspond to onsets. Note that, unlike the DPD-test, such selection does not preclude cases where multiple coherent onsets are present. The method, which we will call hierarchical grid refinement (HiGRID) henceforth, is applied using the SHD coefficients of the selected time-frequency bins. The output of HiGRID for each time-frequency bin is an SRPD map defined on a sparse, multiresolution grid. Segmentation and labelling of this map will jointly provide the number of sources and their DOAs for the analysed bin. This is achieved by using a variant of connected components labelling (CCL) followed by the calculation of cluster centroids on the labelled SRPD map segments. A global histogram of the centroids obtained from all the processed TF bins is then used to estimate the source DOAs. The diagram in Fig.~\ref{fig:flow} shows the proposed algorithm which consists of: 1) pre-processing, 2) SRPD map calculation via grid refinement, 3) local DOA estimation, and 4) post-processing stages. These stages will be explained in more detail in the following sections.

\subsection{Pre-processing: STFT, SHD, and Onset Detection}
Pre-processing involves obtaining the time-frequency representations of the microphone array signals, selecting the time-frequency bins over which DOA estimations will be made, and calculating the SHD coefficients for these bins.

In the first preprocessing stage, a windowed Fourier transform is used to represent the recorded microphone signals in the time-frequency domain, $p_q(\tau, \kappa)$, where $q$ is the index of the microphone, and $\tau$ and $\kappa$ are the time and frequency indices, respectively.

The second preprocessing stage involves the selection of bins at which the proposed algorithm will be applied. For this purpose, spectrum based onset detection is applied on the omnidirectional component, 
\begin{equation}
p_{o}(\tau, \kappa) \approx \frac{1}{Q}\sum_{q=1}^{Q}p_q(\tau, \kappa),
\end{equation}
in order to obtain the time indices, $\{\hat{\tau}\}$, at which the onsets occur. This is followed by the selection of the bins, $(\hat{\tau}, \hat{\kappa})$ whose normalized energy is above a threshold. While many different onset detection algorithms could be suitable for this purpose, we use the \textit{SuperFlux} algorithm that is robust to short-time frequency modulations~\cite{bock2013maximum}. The set of bins that are selected this way will be represented as $\mathcal{T}\triangleq\{(\hat{\tau}, \hat{\kappa})\}$.

The selection of time-frequency bins based on onset detection is not a strictly necessary step for localizing multiple sources. However, it is done for reasons based on the presupposition that sound sources of interest such as speech and music will always have onsets. These reasons are as follows\footnote{While onset detection was chosen for its simplicity, it is not the only way to select TF bins. The proposed method can almost trivially be adapted for use with other TF bin selection methods such as the DPD test~\cite{nadiri2014localization} or sound field directivity test~\cite{Rafaely:2017cj}.}: 
\begin{enumerate}
\item Since the direct path from \textit{at least one} of the sources will always be present at the onset time frames, selecting these frames for analysis would improve estimation accuracy, and 
\item The number of time-frequency bins to be processed, and thus the associated computational cost would be significantly reduced.  
\end{enumerate}

In the third pre-processing stage, the SHD coefficients are calculated as in (\ref{eq:DSHD}) such that:
\begin{equation}
p_{nm}(\hat{\tau}, \hat{\kappa})=\sum_{q=1}^{Q}w_qp_q(\hat{\tau}, \hat{\kappa})[Y_n^m(\theta_q,\phi_q)]^*.
\end{equation}
for all the selected bins, $(\hat{\tau}, \hat{\kappa})\in{}\mathcal{T}$. These coefficients are used in the calculation of SRPD maps at their respective time-frequency locale.

\subsection{Hierarchical Grid Refinement (HiGRID)}
The proposed algorithm consists of the identification and refinement of  regions of interest in the SRPD map that contain sources, followed by the clustering of contiguous regions to count and localize these sources. 

\subsubsection{Analysis Grid}
The selection of a suitable spherical analysis grid is essential. A necessary property is that it is hierarchical such that each refinement of the grid also contains the coarser grid and also each grid element can be represented as a node in a tree structure. A desirable property is that the grid is approximately, even if not exactly, uniform.

Hierarchical Equal Area isoLatitude Pixelization (HEALPix)~\cite{Gorski:2005ku} is used here as it allows an efficient representation of spherical data using quadrilateral, non-overlapping, and refinable grid elements (i.e. pixels) on a sphere. At the resolution level, $l\in\mathbb{N}$, HEALPix provides spherical tessellation with $12\cdot2^{2l}$ equal area grid elements. The elements are separated uniformly with an angular resolution (in radians) of:
\begin{equation}
\Theta_\Delta = \sqrt{\frac{3}{\pi}}\frac{\pi}{3\cdot{}2^{l}},
\end{equation}
and their areas depend on the resolution level such that:
\begin{equation}
A_l= \frac{4\pi R^2}{12\cdot{}2^{2l}},
\end{equation} 
where $R$ is the radius of the sphere~\cite{Gorski:2005ku}. Note that at the lowest resolution level (i.e. $l=0$) there are $12$ grid elements. 

Mesh refinement partitions a grid element, $\mathcal{S}_{l,m}$ at the resolution level $l$ with index $m$, into four new grid elements neighboring each other, at the higher resolution level, $l+1$, such that:
\begin{equation}
\mathcal{S}_{l,m}=\mathcal{S}_{l+1,4m}\oplus\mathcal{S}_{l+1,4m+1}\oplus\mathcal{S}_{l+1,4m+2}\oplus\mathcal{S}_{l+1,4m+3}
\end{equation}
where $\left\{\mathcal{S}_{(l+1,4m+k)}\right\}$ for $k=0\cdots{}3$ are four higher resolution grid elements. This allows for a \textit{quadtree} representation  where a spherical function or distribution can be represented at different levels of detail for different directions. The flexible multiresolution representation afforded by HEALPix is essential for the DOA estimation and source counting method proposed in this paper.

\subsubsection{Entropy-based Hierarchical Grid Refinement}
When multiple coherent sources are present in the sound scene, SRPD maps will have more than one meaningful local maximum, that correspond to the directions of the active sources. In order to find these maxima without steering a beam in all possible directions, it is necessary to identify the regions that contain these maxima, and scan these regions with a higher resolution. Spatial entropy~\cite{batty1974spatial}, which is a measure of the spatial disorganization, is used to identify such regions. 

Let us assume that at a given iteration, $t$, the analysis grid (or equivalently the leaf nodes of the corresponding quadtree), $\mathcal{G}_t$ for a time frequency bin $(\hat\tau, \hat\kappa)$ consists of $L_t$ elements (i.e. nodes) $\mathcal{G}_t=\{\mathcal{S}_{l,m}^{(t)}\}$ at different resolution levels, $l\in\{0\cdots{}t-1\}$.  We can define the total spatial entropy of the representation as:
\begin{equation}
H(\mathcal{G}_t)=-\sum_{\forall{}\mathcal{S}_{l,m}^{(t)}\in\mathcal{G}_t}\gamma(\mathcal{S}_{l,m}^{(t)})\log{}\frac{\gamma(\mathcal{S}_{l,m}^{(t)})}{A(\mathcal{S}_{l,m}^{(t)})}
\end{equation}
and
\begin{equation}
\gamma(\mathcal{S}_{L,M}^{(t)})=\frac{\mathcal{P}_{L,M}}{\sum_{\forall{}\mathcal{S}_{l,m}^{(t)}\in\mathcal{G}_t}\mathcal{P}_{l,m}}\label{eq:gamma}
\end{equation}
where $\mathcal{P}_{l,m}$ is the SRPD of the grid element with index $m$ at level $l$ [as in (\ref{eq:srpdwm})], and $A(\mathcal{S}_{l,m}^{(t)})$ is its area. Here, $0\leq\gamma(\mathcal{S}_{l,m}^{(t)})\leq{}1$ is interpreted as the probability of the grid element $\mathcal{S}_{l,m}^{(t)}$ of containing a source, and:
\begin{equation}
\sum_{\forall{}\mathcal{S}_{l,m}^{(t)}\in\mathcal{G}_t}\gamma(\mathcal{S}_{l,m}^{(t)})=1.
\end{equation}

The decision on whether or not to refine a specific grid element is made according to its effect on total spatial entropy. A candidate grid $\mathcal{G}_t'$ can be obtained from the existing grid $\mathcal{G}_t$ by refining a specific grid element, $\mathcal{S}_{l,m}^{(t)}$ such that:
\begin{equation}
\mathcal{G}_{t}' = \mathcal{G}_{t} \cup \mathcal{V}(\mathcal{S}_{l,m}) \setminus \{\mathcal{S}_{l,m}^{(t)}\}.
\end{equation}
where $\mathcal{V}(\mathcal{S}_{l,m})=\{\mathcal{S}_{l+1,4m}, \mathcal{S}_{l+1,4m+1}, \mathcal{S}_{l+1,4m+2}, \mathcal{S}_{l+1,4m+3}\}$ 
is the set of children nodes in the quadtree representation of the analysis grid.

The proposed algorithm aims to decrease the total spatial entropy of the representation by using the difference between the total entropy before and after the refinement, i.e.~: 
\begin{equation}
\mathcal{I}(l,m)=H(\mathcal{G}_t)-H(\mathcal{G}_t').
\end{equation}
This difference is known as \textit{mutual information} or \textit{information gain} in the context of decision trees~\cite{murphy2012machine} and, when positive, indicates that the refinement would result in the creation of children nodes with similar SRPD values indicating a local maximum. Therefore, if the refinement decreases the total spatial entropy such that, $\mathcal{I}(l,m)>0$, the representation is updated with the refined candidate grid, $\mathcal{G}_t'$. Otherwise the corresponding branch of the quadtree is pruned at its present resolution level and removed from the search path. Algorithm \ref{algo:higrid} shows the pseudocode for the HiGRID method.


\begin{figure}[!t]
 \removelatexerror
  \begin{algorithm}[H]\label{algo:higrid}
   \caption{Hierarchical Grid Refinement (HiGRID)}
   \KwData{SHD coefficients, $p_{nm}$ for the time-frequency bin $(\hat{\tau}, \hat{\kappa})$ and the maximum refinement level, $L_{\max}$}
   \KwResult{HEALPix quadtree with the set of leaf nodes, $B=\{\mathcal{S}_{l,m},\ l=0..L\}$ containing the SRPD evaluated at the corresponding pixel, $\mathcal{P}_{l,m}$}
    
    $treeLevel \leftarrow{}1$
    
    Initialize the quadtree by calculating the SRPD at each leaf node, $\mathcal{P}_{1,m}$
   
        $N\leftarrow{}\emptyset$
        
        \While{treeLevel $\leq{}L_{\max}$}
        {
        	$B=\{\mathcal{S}_{l,m}\ |\ l=treeLevel\}$
        	
            \While{$B\neq\emptyset$}
            {
            
            $\mathcal{S}\leftarrow$fetchRandomLeafNodeFrom($B$)
            
            $C\leftarrow{}$childrenOf($\mathcal{S}$) \tcp*[h]{Calculate SRPDs for $C=\{C_i, i=1\cdots{}4\}$}
            
            $B'=B \cup C \setminus S$
            
            $E_B\leftarrow{}spatialEntropy(B)$
            
            $E_{B'}\leftarrow{}spatialEntropy(B')$
            
            \eIf{$E_{B'}<E_B$}
                {
                $N\leftarrow{}N\ \cup\ C$
                }
                {
                $N\leftarrow{}N\ \cup\ \{\mathcal{S}\}$
                }
            $B\leftarrow{}B\ \setminus \ \mathcal{S}$
            }
        $B\leftarrow{}N$
        
        $treeLevel \leftarrow{} treeLevel + 1$
        }
  \end{algorithm}
\end{figure}

\begin{figure}[!t]
\centering
\subfigure[]{\includegraphics[width=.24\textwidth]{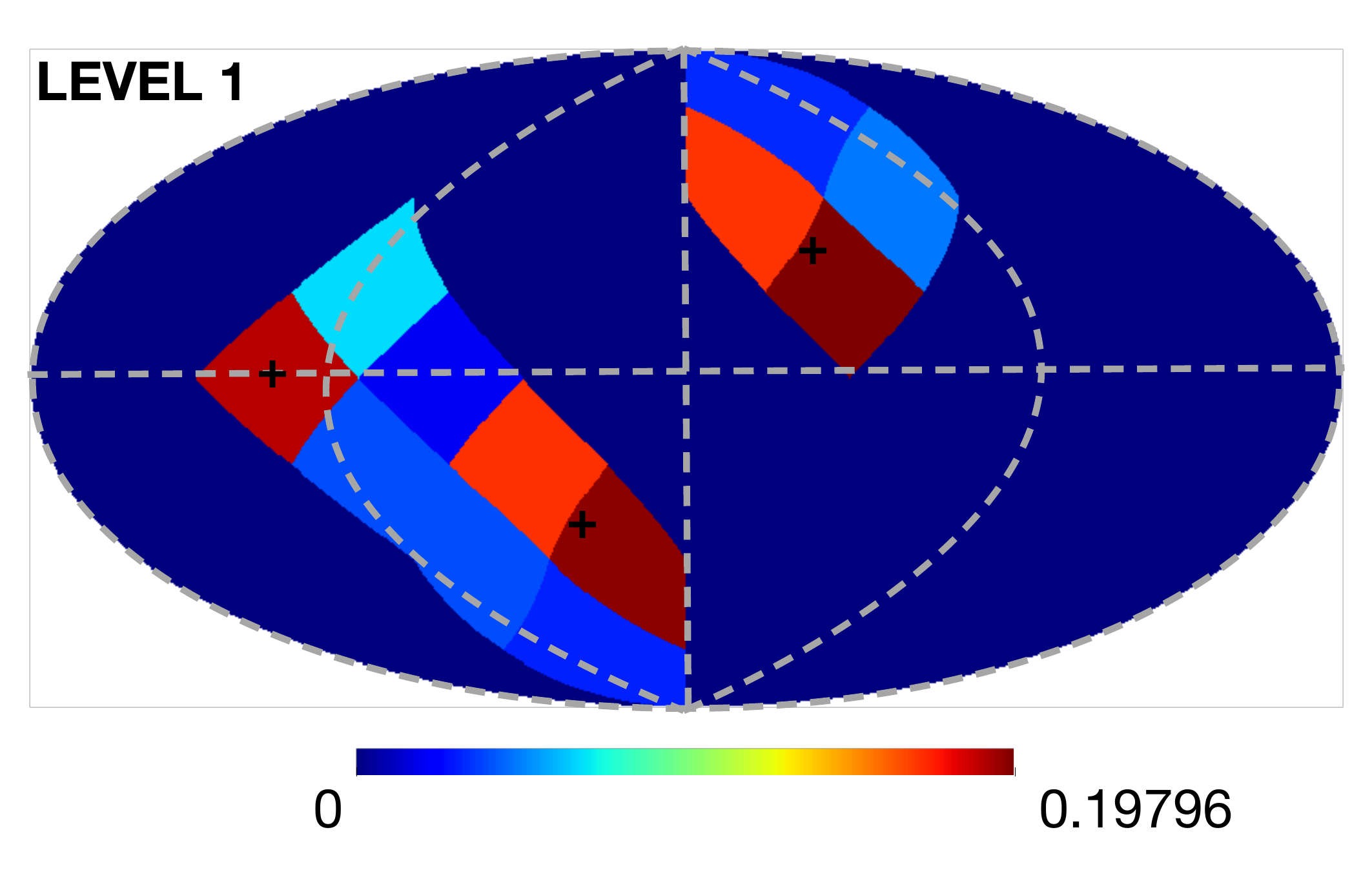}\label{fig:lev1}}
\subfigure[]{\includegraphics[width=.24\textwidth]{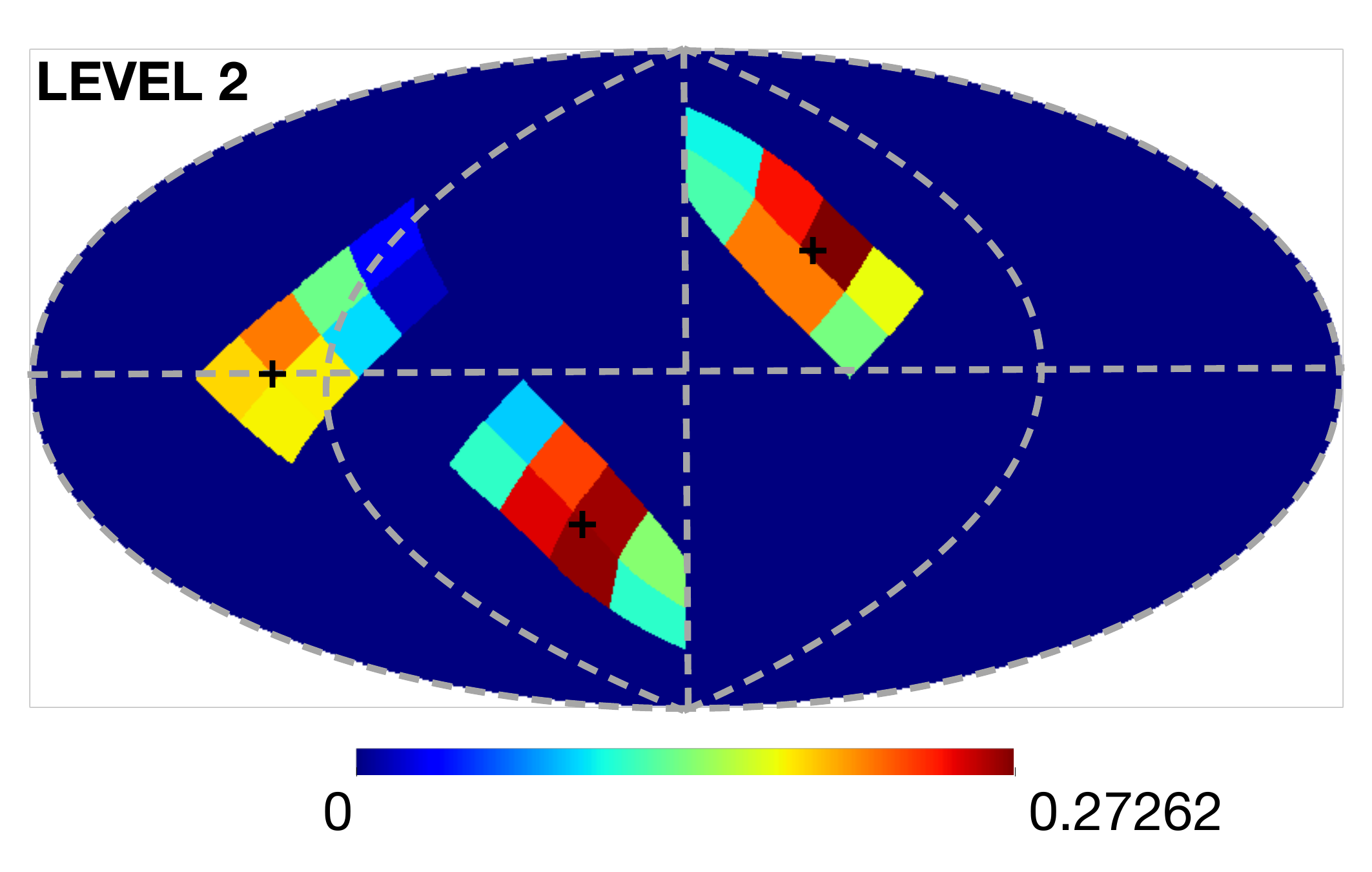}\label{fig:lev2}}
\subfigure[]{\includegraphics[width=.24\textwidth]{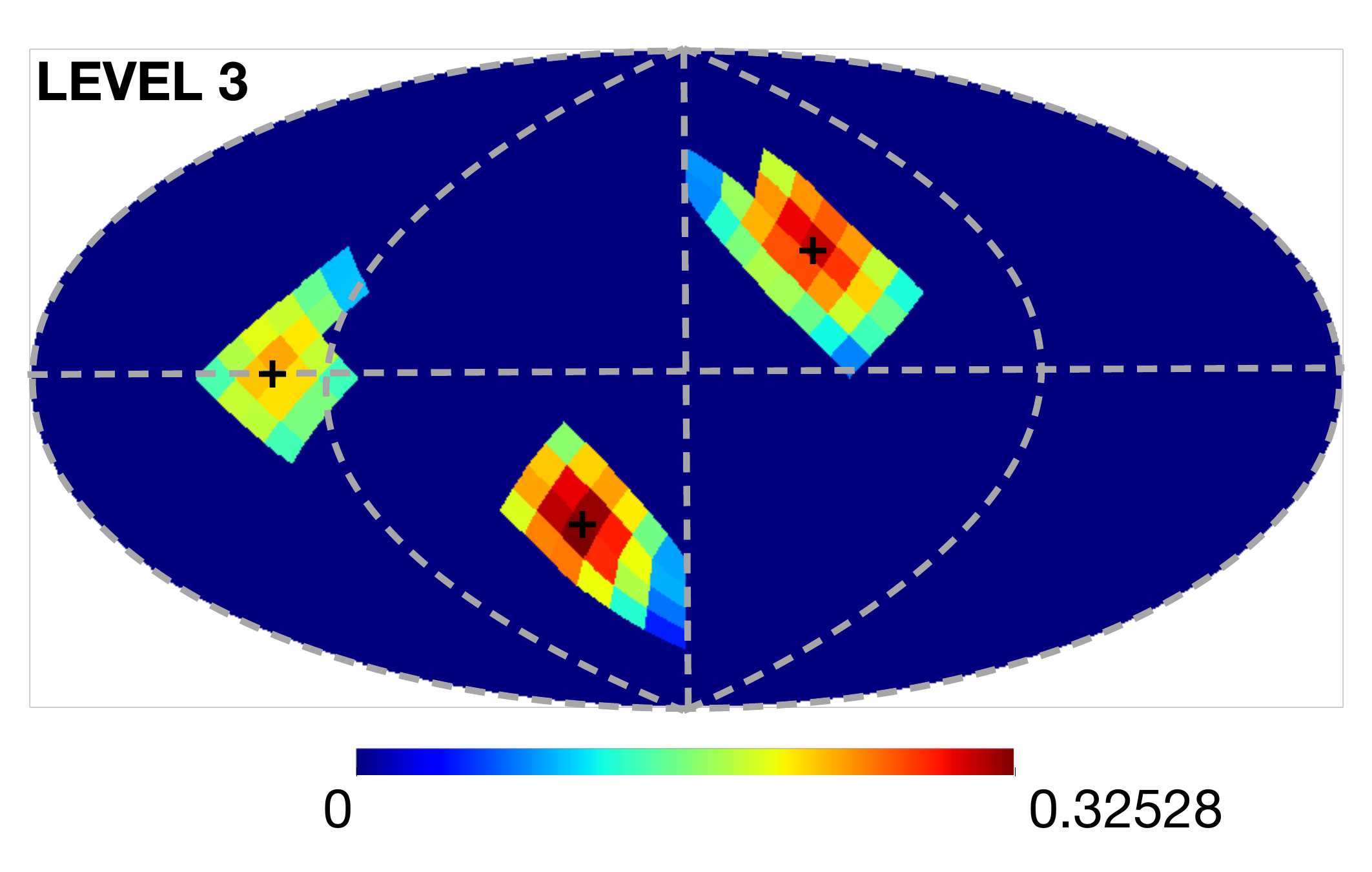}\label{fig:lev3}}
\subfigure[]{\includegraphics[width=.24\textwidth]{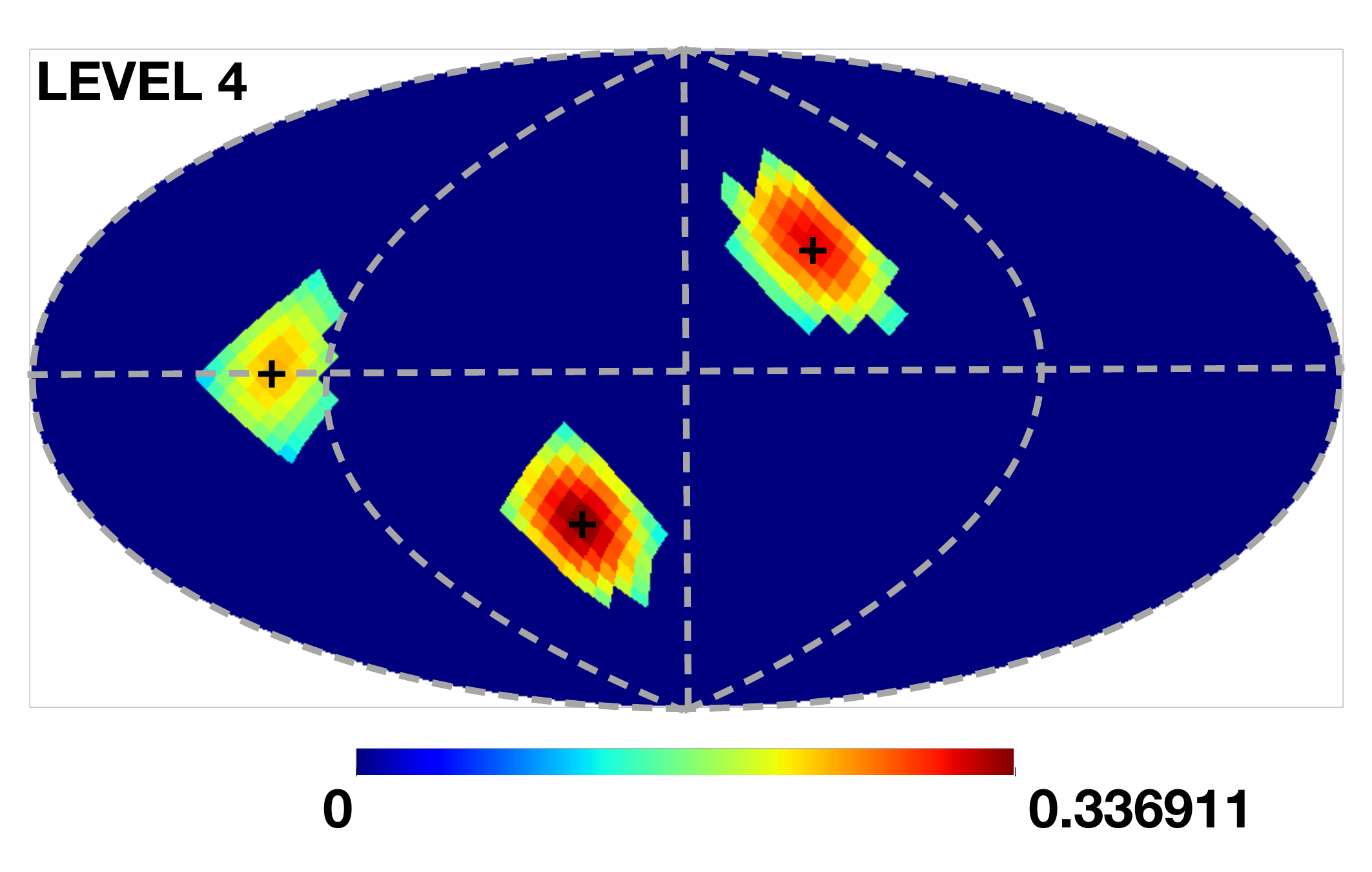}\label{fig:lev4}}
\caption{Hiearchical grid refinement (HiGRID) applied on a single time-frequency bin, representing a sum of three unit amplitude, monochromatic plane waves with a frequency of $F=3$ kHz. The progress of the algorithm at levels 1-4 are shown in (a)-(d), respectively. }\label{fig:pwdecom}
\end{figure}

Fig.~\ref{fig:pwdecom} shows the progress of the proposed algorithm using the SHD coefficients of the sum of three unit amplitude, monochromatic plane waves with the common frequency of $F=3$ kHz, incident on a spherical aperture of radius $4.2$ cm from the directions of $(\phi_1, \theta_1)=(\pi/2, 3\pi/5)$, $(\phi_2, \theta_2)=(2\pi/3, \pi/5)$, and $(\phi_3, \theta_3)=(\pi/3, 9\pi/5)$, respectively. The real DOAs of the plane waves are denoted by black cross signs. The decomposition at different resolution levels, from 1 through to 4, obtained using the HiGRID algorithm are shown using the Mollweide projection. It may be observed that the SRPD map contracts in regions corresponding to the DOA of each plane wave. SRPD is calculated at 33, 72, 165 and 429 possible directions for levels 1, 2, 3, and 4, respectively. In order to attain the same angular resolution without entropy-based selection of regions of interest, 48, 192, 768 and 3072 directions would have had to be scanned at each of these different levels.

\subsubsection{Source Localization}
Interpreting the HEALPix grid as a quadtree allows a natural clustering of the underlying data that the grid represents. Once the multiresolution SRPD map is  formed up to a given resolution level, the corresponding quadtree can provide a spatial map of the recorded scene with the leaves (i.e.~pixels; grid elements) at the highest level representing the probable source directions. We adapt the \textit{connected components labelling} (CCL)~\cite{soille2013morphological} technique from the computer vision domain for use with the quadtree representation of the grid to identify and label regions of interest based on neighbourship relations between the nodes. The algorithm which we call \textit{neighbouring nodes labelling} (NNL) is shown in Algorithm~\ref{algo:nnl}\footnote{It should be noted that NNL algorithm is provided here for convenience and not for its novelty since there exist similar applications of CCL in quadtree representations of data e.g.\ \cite{samet1981connected}.}.

\begin{figure}[!t]
 \removelatexerror
  \begin{algorithm}[H]\label{algo:nnl}
   \caption{Neighboring Nodes Labelling (NNL)}
   \KwData{Set of leaf nodes, $\mathcal{G}=\{\mathcal{S}_{l,m}\ l=0..L\}$ of a HEALPix quadtree containing the values of SRPD distribution, $\mathcal{P}_{l,m}$}
   \KwResult{Sets of nodes $C_s$ where $s=1\cdots{}S$ }
   \tcp*[h]{Binarization via thresholding}
   $THR\leftarrow$calculateThreshold($\{\mathcal{P}_{l,m}\}$)

   \For{$\{\mathcal{S}_{l,m}\}\in{}\mathcal{G}$}{
       \If{$\mathcal{P}_{l,m}<THR$}{Remove node $\mathcal{S}_{l,m}$ from $\mathcal{G}$}
       }
           
    $s\leftarrow{}1$     \tcp*[h]{Initialize the label count}
    
    $\mathcal{G}_L = \{\mathcal{S}_{l,m}\ |\ l=L_{\max}\}$ \tcp*[h]{Select leaf nodes at the highest resolution level}
    
    \While{$\mathcal{G}_L \neq \emptyset$}{
        stopFlag$\leftarrow$\texttt{True}
        
        $C_s\leftarrow\emptyset$ \tcp*[h]{Initialize cluster s}
        
        $\mathcal{S}\leftarrow$fetchRandomElementFrom($\mathcal{G}_L$)
        
        $C_s\leftarrow{}C_s\cup\{\mathcal{S}\}$
        
        $\mathcal{G}_L\leftarrow{}\mathcal{G}_L\setminus{}\{\mathcal{S}\}$

       \While{stopFlag is \texttt{True}}{
           $N\leftarrow$neighborsOfSet($C_s$)
           
           $D\leftarrow{}\mathcal{G}_L\cap{}N$
           
           \eIf{$D\neq{}\emptyset$}{
               $C_s\leftarrow{}C_s\cup{}D$
               
               $\mathcal{G}_L\leftarrow{}\mathcal{G}_L\setminus{}D$ 
           }
           {
            $s\leftarrow{}s+1$
            
            stopFlag$\leftarrow$\texttt{False}
            
            }
       }

    }
    
  \end{algorithm}
\end{figure}

CCL and NNL are similar in the way that they identify regions based on their contiguity and that they carry out labelling and clustering jointly. The major differences between CCL and NNL are that NNL can operate on a multiresolution representation whereas CCL operates on uniformly sampled pixels. Also, as opposed to CCL, NNL takes into account the periodic boundaries (i.e. at $\phi=2\pi$, and $\theta=0, \pi$) on the spherical grid while identifying and labelling contiguous regions. 

Similar to CCL which operates on binarized images, NNL operates on a set of leaf nodes whose values are above a threshold. Thresholding also allows obtaining more compact regions of interest. While it is possible to employ more elaborate thresholding techniques such as Bradley's adaptive thresholding~\cite{bradley2007adaptive}, we used the mean value of the leaf nodes of the tree as a threshold due to its simplicity.

Once contiguous regions are identified, the centroid of each region, $\mathbf{n}(\hat{\tau}, \hat{\kappa})$, is stored as the local DOA estimation for one of the sources in the time-frequency bin. Note that the directions of these unit vectors will not necessarily coincide with the centres of HEALPix grid elements.

\subsection{Post-processing}
Post-processing stage uses a 2D direction histogram (with a uniform $1^{\circ}$ bin size for $\theta$ and $\phi$) obtained from the set of unit vectors denoting source DOAs, $\{\mathbf{n}(\hat{\tau}, \hat{\kappa})\}$. Three operations are used on this histogram prior to the final DOA estimation:
\begin{enumerate}
\item Due to the small number of DOA estimates from the selected time-frequency bins, the directional histogram will be sparse. In order to denoise the 2D histogram, spurious bins with a single occurrence are eliminated and a median filter with a neighbourhood size of 3 is applied to the resulting 2D histogram.

\item The denoised histogram is then processed with a Gaussian filter with a kernel size of 3 to obtain a dense representation that is suitable for segmentation-based clustering.

\item The resulting histogram is then cast back to a HEALPix grid to be used with the NNL method described earlier. Finally, NNL is applied on this aggregate representation and centroids of the identified clusters are registered as source DOAs.
\end{enumerate}

\section{COMPUTATIONAL COST}\label{sec:comp}
It was shown in the previous section that using HiGRID, it is possible to obtain DOA estimates using far fewer steering directions than SRP. However, there is some overhead due to the calculation of information gain and also due to the calculation of SRPD which is different than SRP. An analytic derivation of the asymptotic complexity of HiGRID is very hard if not impossible to obtain. Therefore, the  time required to process a single time-frequency bin is used as a measure to compare the computational performances of HiGRID and SRP.

The comparisons that we present here involved simulating up to 18 monochromatic plane waves ($F=3$ kHz), incident from random directions with at least $\pi/4$ separation. A number of 50 repetitions were made for each source count. Vectorised implementations of both methods programmed in Python 2.7, running on a MacBook Pro with Intel Core i5 2.9 GHz and 16 GB RAM were used.

\begin{figure}[!t]
\centering
\includegraphics[width=.5\textwidth]{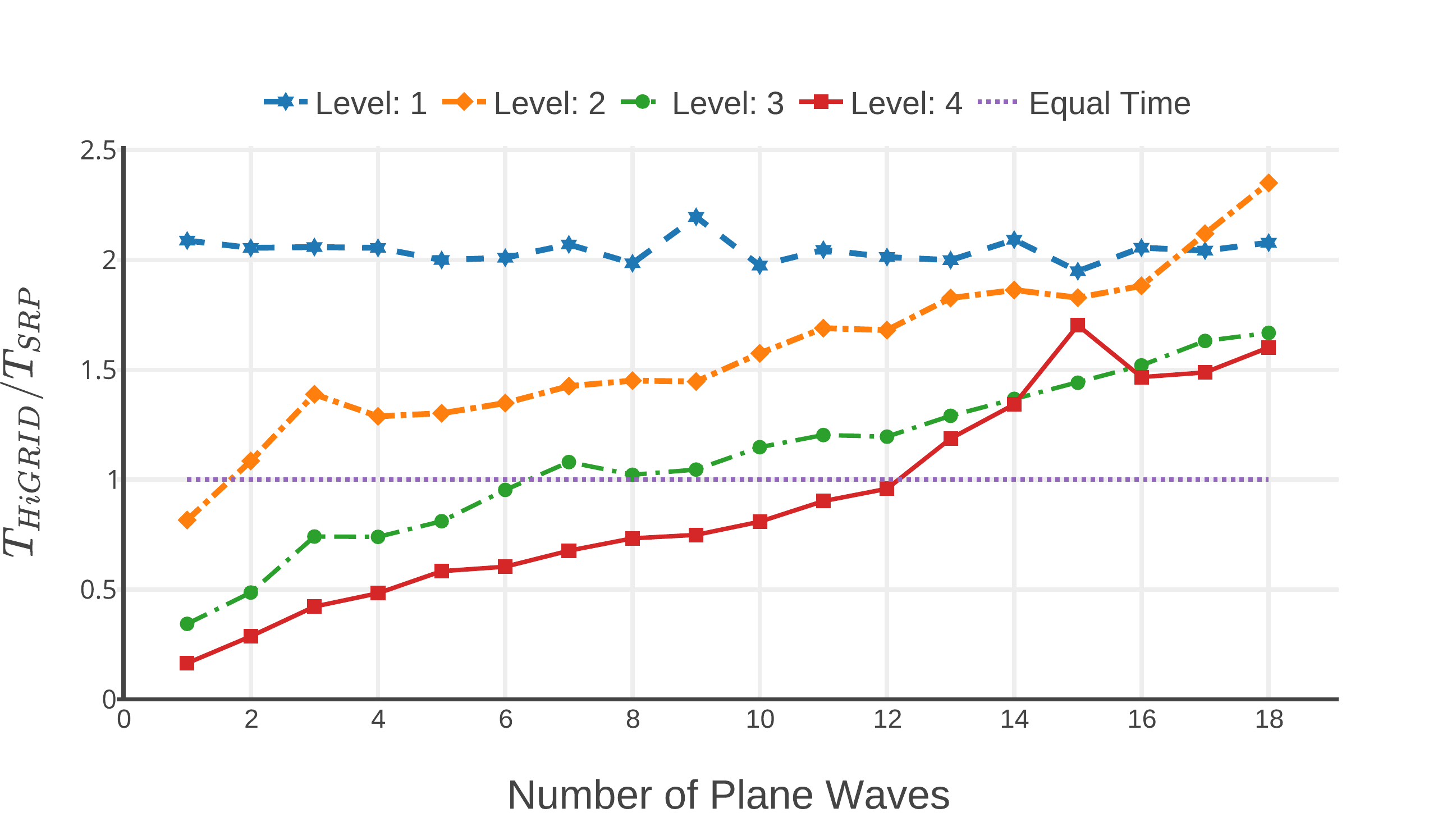}
\caption{Ratio of computing time of HiGRID ($T_{HiGRID}$) to SRP ($T_{SRP}$) for different resolution levels from 1 to 4 and for different number of coherent plane waves with random directions. Each data point represents the ratio calculated over the average of computing times for 50 simulations.}\label{fig:comptime1}
\end{figure}

\begin{figure}[!t]
\centering
\includegraphics[width=.5\textwidth]{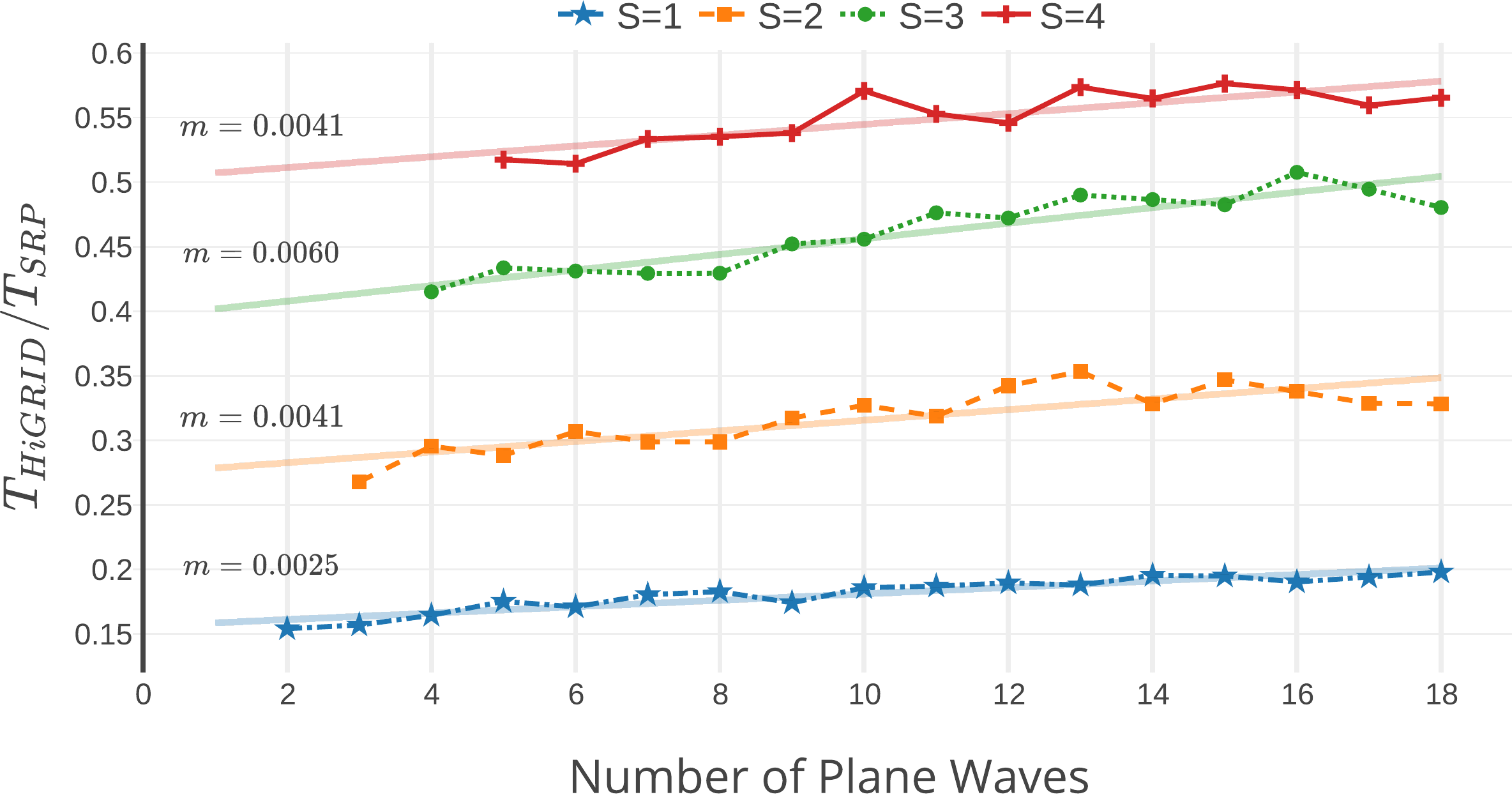}
\caption{Ratio of computing time of HiGRID ($T_{HiGRID}$) to SRP ($T_{SRP}$) for a total of 18 plane waves with different number of, $S=1$, $S=2$, $S=3$ and $S=4$ coherent plane waves with random directions. Each data point represents the ratio calculated over the average of computing times for 50 simulations.}\label{fig:comptime2}
\end{figure}

Fig.~\ref{fig:comptime1} shows the ratio of the time spent using HiGRID ($T_{HiGRID}$) to time spent using SRP ($T_{SRP}$) for different numbers of coherent, unit amplitude plane waves. It may be observed that HiGRID provides drastic computational advantages for higher resolution (e.g. at level 4) and at lower source counts in comparison with SRP. At the resolution level 4 and for a single unit amplitude plane wave, computation time required for HiGRID is $16.4\%$ of that required by SRP. The computation time becomes similar for $12$ plane waves. At level 3, the computational time is similar for $6$ plane waves. It may be observed that at level 1, SRP outperforms HiGRID in terms of computation time by a factor of 2. This is due to the dimensions of the basis used in SRPD calculations. However, since the angular resolution at this level is $29.3^{\circ}$, which is too low to be useful in a practical application, this difference is not relevant. Note that cases with many unit amplitude plane waves are not representative of real life scenarios as it is unlikely that a high number of fully coherent plane waves with equal amplitudes will impinge on the array simultaneously. However, these conditions represent the worst-case scenarios and hence chosen to demonstrate the computational efficiency of the proposed algorithm in comparison with SRP. 

In order to evaluate the computational performance of the proposed method in a more realistic scenario, simulations of a total of up to 18 plane waves were carried out at a resolution level of 4. In the four different simulations, $S=1$, $S=2$, $S=3$, and $S=4$ were unit amplitude, coherent plane waves, respectively. The remaining plane waves were assigned random, complex-valued amplitudes with magnitudes drawn from a uniform distribution in $[0,0.5]$. This represents a situation similar to a combination of direct and diffuse fields. Fig.~\ref{fig:comptime2} shows the ratio of the time spent using SRP to time spent using HiGRID as well as lines fit to the averages of these ratios for increasing number of components. It may be observed that the ratio is below $20\%$, $35\%$, $51\%$ and $57\%$ for $S=1$, $S=2$, $S=3$, and $S=4$ coherent plane waves, respectively. It increases linearly with the total number of components. However, this trend is very slow with the maximum slope of $0.6\%$ for $S=3$.

\section{PERFORMANCE EVALUATION}\label{sec:eval}
The proposed algorithm is evaluated using acoustic impulse responses (AIRs) measured in a highly reverberant classroom. Robustness of the proposed method to additive white noise is assessed via emulated acoustic scenes. A separate comparison with the state-of-the-art methods is presented. In addition a microphone array recording of a musical performance was used to demonstrate the performance of the algorithm in a real-life scenario.

\subsection{Acoustic Impulse Response Measurements}
In order to provide flexibility in terms of creating different test scenarios, we measured multichannel acoustic impulse responses (AIRs) in an empty classroom at METU Graduate School of Informatics with a rigid spherical microphone array (mh Acoustics Eigenmike em32). Logarithmic sine sweep method~\cite{farina2000simultaneous} was used in the measurements. The sound source was a Genelec 6010A two-way loudspeaker whose acoustic axis pointed at the vertical axis of the array at all measurement positions. Eigenmike em32 has 32 microphones positioned on a rigid sphere on the vertices of a truncated icosahedron and allows the decomposition of the sound field up to fourth order (i.e.~$N=4$)~\cite{meyer2004spherical}.

\begin{figure}[!t]
\centering
\includegraphics[width=.5\textwidth]{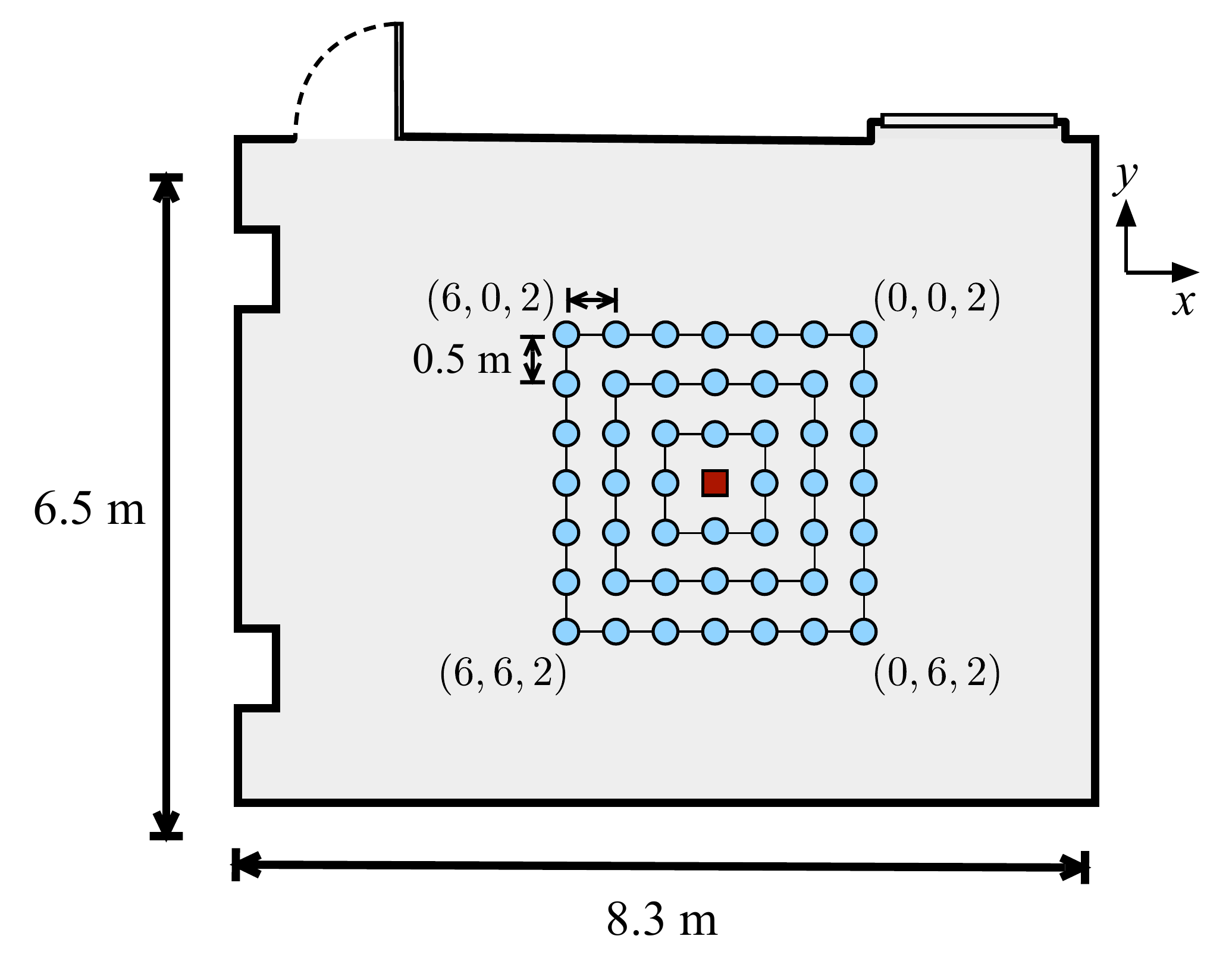}

\caption{Top view of the measurement positions inside the classroom. Circles represent the source locations and the square represents the position of the microphone array. The walls are made of concrete covered with plaster, the floor is carpeted, and the ceiling has acoustic tiles. One of the walls also has a window and a wooden door that were both kept closed during the measurements.}\label{fig:room}
\end{figure}

The classroom in which the measurements were made has a high reverberation time ($T_{60}\approx{}1.12$ s) when empty. It is approximately rectangular and has the dimensions $6.5\times{}8.3\times{}2.9$ m. AIR measurements were made at 240 points on a rectilinear grid of $0.5$ m horizontal and $0.3$ m vertical resolution surrounding the array. The array was positioned at a height of $1.5$ m. The measurement planes were positioned at the heights of $0.9$, $1.2$, $1.5$, $1.8$ and $2.1$ m from the floor level. These positions cover the whole azimuth range and an elevation range of approximately $\pm{}50^{\circ}$ above and below the horizontal plane. We also measured room impulse responses (RIRs) at the same positions using an Alctron M6 omnidirectional microphone. This second set of measurements were used to calculate the corresponding direct-to-reverberant (D/R) energy ratios for these positions. Fig.~\ref{fig:room} shows the top view of the measurement positions. 

\subsection{Source signals}
In order to evaluate the proposed algorithm with near-incoherent (i.e. nominally W-disjoint) sources, we used the first 4 seconds of anechoic speech recordings from B\&{}O Music for Archimedes CD~\cite{archimedes}. These signals included speech recordings in English and Danish from two female and two male speakers. These were resampled to $48$ kHz. 

\begin{figure}
\centering
\includegraphics[width=.5\textwidth]{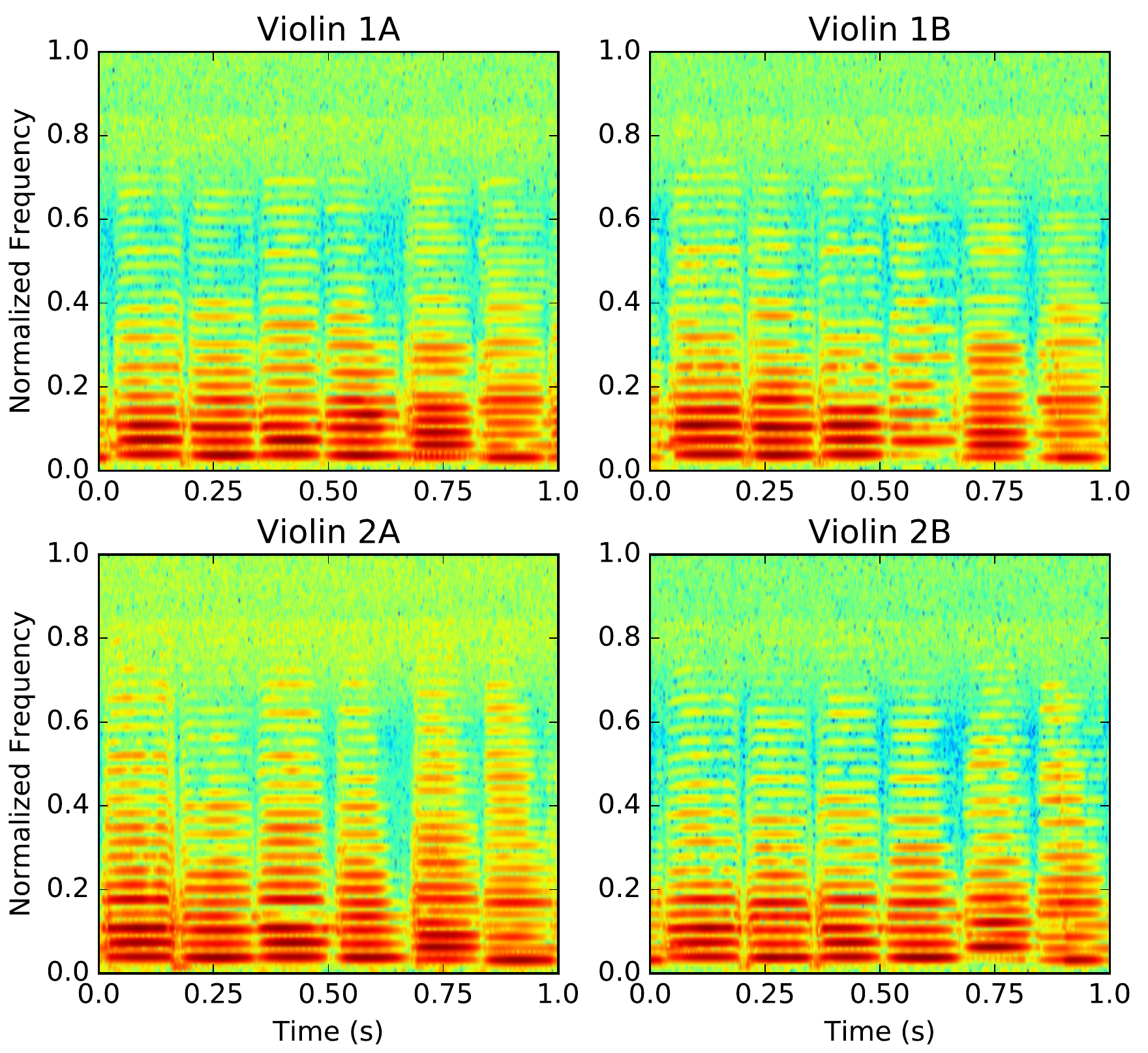}
\caption{Magnitude spectrograms of the first second of the four violin tracks used in the evaluations.}\label{fig:violins}
\end{figure}

In order to evaluate the proposed algorithm with near-coherent sources, we used 4 seconds (01:00-01:04) of the anechoic recordings of Gustav Mahler's Symphony Nr. 1, fourth movement~\cite{patynen2008anechoic}. Only the four violin tracks were used in which violins play the same musical phrase in unison. It should be noted that while there may be minor differences, the source signals have substantially similar magnitude spectrograms (see Fig.~\ref{fig:violins}). This selection of test signals constitutes a very exacting, albeit realistic scenario, for example in object-based audio capture. The sampling rate was $48$ kHz.

The speech signals were normalized according to ITU P.56~\cite{ITUP56} using the \texttt{VOICEBOX} toolbox~\cite{brookes1997voicebox}. The violin signals were normalized for energy.

\subsection{Performance Indicators}
We use three indicators for assessing the performance of DOA estimation performance. The first indicator is the average DOA estimation error defined as:
\begin{equation}
\epsilon_{\textrm{DOA}}=\frac{1}{S_{est}}\sum_{s=1}^{S_{est}}|\arccos\langle\mathbf{n}_{\textrm{est,i}},\mathbf{n}_{\textrm{src,i}}\rangle|
\end{equation}
where $S_{est}$ is the number of correctly estimated sources, $\mathbf{n}_{\textrm{est,i}}$ and $\mathbf{n}_{\textrm{src,i}}$ are the unit vectors in the direction of the estimated DOA and the nominal source direction, respectively. The closer this value to zero, the better the DOA estimation.

The second performance indicator is the average number of localised sources, $S_{\textrm{avg}}$. While this indicator may provide insight about the average success of the algorithm in identifying the source count, it may also be misleading. For example, if for a scenario involving 4 sources, 8 sources are identified in one trial and none in another trial, the average would indicate ideal performance.

The third indicator depends on the source count disparity defined as:
\begin{equation}
\Delta_s =S_{est}-S_{act}
\end{equation}
where $S_{act}$ is the actual number of sources. Ideally, $\Delta_s=0$ (i.e.\ $S_{est}=S_{act}$). It should be noted that $\Delta_s<0$ (e.g. $S_{est}<S_{act}$) is a more serious problem than $\Delta_s>0$ (e.g. $S_{est}>S_{act}$) since the former indicates one or more undetected sources, while the latter may indicate that all sources in the scene are detected alongside some strong coherent reflections. Therefore, the third performance indicator, \textit{negative source disparity}, is defined as the number of times the source count disparity is negative in a given number of trials.

\subsection{Evaluation of the HiGRID algorithm}

In the performance tests explained below, HiGRID algorithm was used with a maximum tree level of $3$ which corresponds to a grid resolution of $7.33^{\circ}$. The maximum order of SHD was $N=4$. The frequency range was between $2608$ Hz and $5216$ Hz for which the mode strength for order, $N=4$, is high. The FFT size and the hop size were $1024$ and $64$ samples, respectively. 

For each of the test cases, the selection of the prescribed number of sources is made randomly, subject to two criteria. The source positions were first grouped according to their D/R ratios. Seven D/R ratio clusters with a cluster width of $1.38$ dB were used. The cluster centres corresponded to $-1.65$, $-0.27$, $1.11$, $2.50$, $3.88$, $5.26$, $6.65$ dB, respectively. Source combinations in each test scenario included positions only from one of these clusters so that the D/R ratio was similar across different source positions. It was also made sure that the separation between any pair of positions in a scenario were greater than $\pi/4$. Ten test scenarios were randomly generated for each of the $S=1$, $2$, $3$, and $4$ source cases and for each of the D/R ratio clusters. Kuhn-Munkres algorithm~\cite{kuhn1955hungarian} was used to assign DOA estimations to the ground truth. In all trials, if the DOA estimation error was found to be larger than $\pi/4$, that case was identified as an extreme value and excluded from the analysis.

\subsubsection{Ideal conditions} 

This condition corresponds to a noiseless scenario where the source signals were convolved with the measured AIRs and summed to obtain the test signals. 

\begin{figure}[!t]
\centering
\includegraphics[width=.5\textwidth]{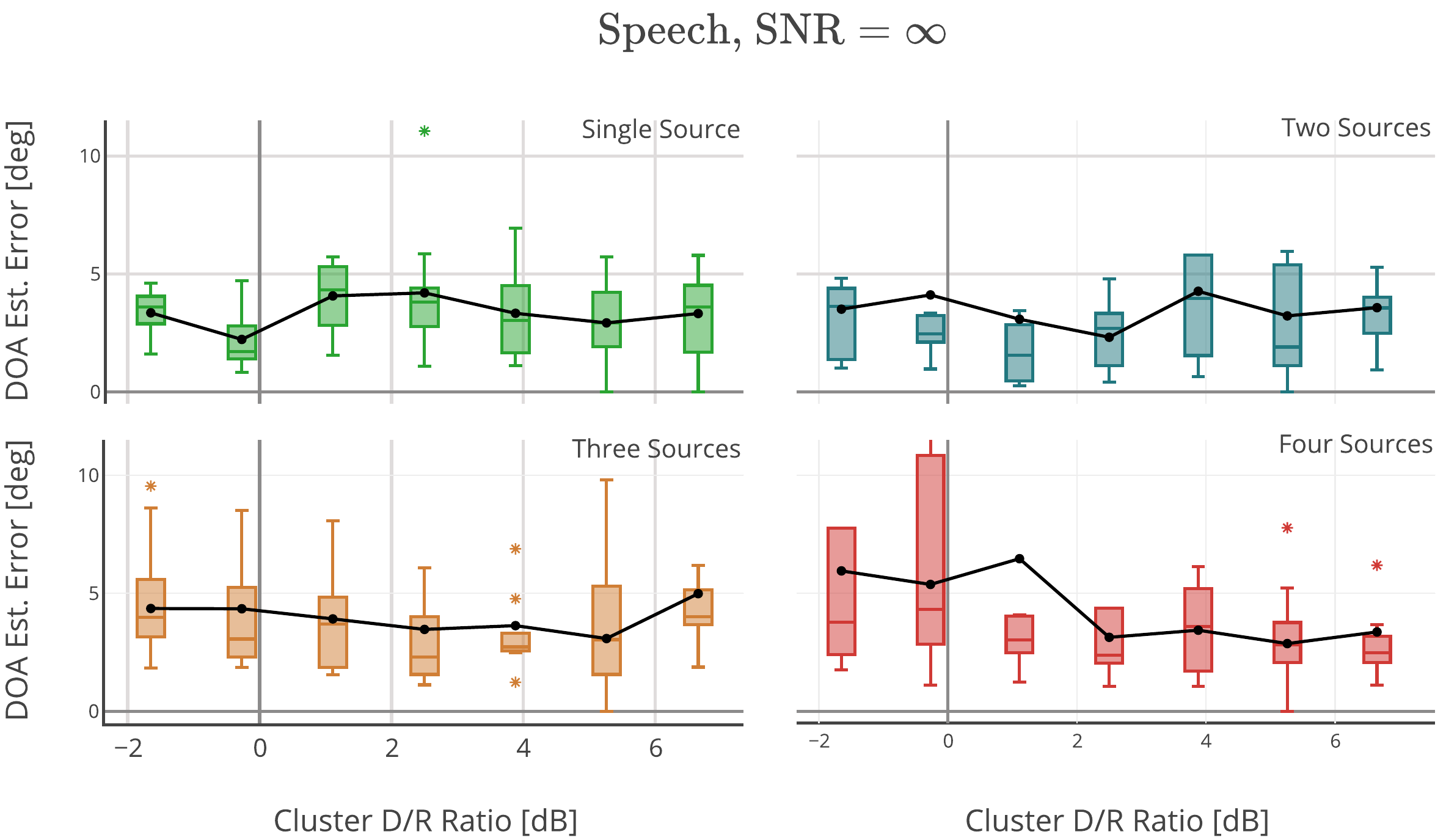}
\caption{Results of DOA estimation for speech signals using HiGRID for $S=1$, $2$, $3$, and $4$ sources and different D/R ratio levels. The box plots indicate the distribution of error. The solid lines indicate the mean DOA error. Asterisks indicate outliers. Extreme values are not shown.}\label{fig:resultsspeech}
\end{figure}

Fig.~\ref{fig:resultsspeech} shows the DOA estimation errors for the case where speech signals are used. There were a total of $19$ extreme values in $280$ distinct test scenarios and $700$ source positions. These were not included in the calculation of the DOA estimation error average. The average DOA estimation errors for $S=1$, $2$, $3$, and $4$ sources were $2.71^{\circ}$, $2.99^{\circ}$, $3.20^{\circ}$, and $3.39^{\circ}$, respectively. The overall average was $3.18^{\circ}$. It may be observed that the DOA estimation error is affected neither by the D/R ratio nor by the number of sources in the scene.

\begin{figure}[!t]
\centering
\includegraphics[width=.5\textwidth]{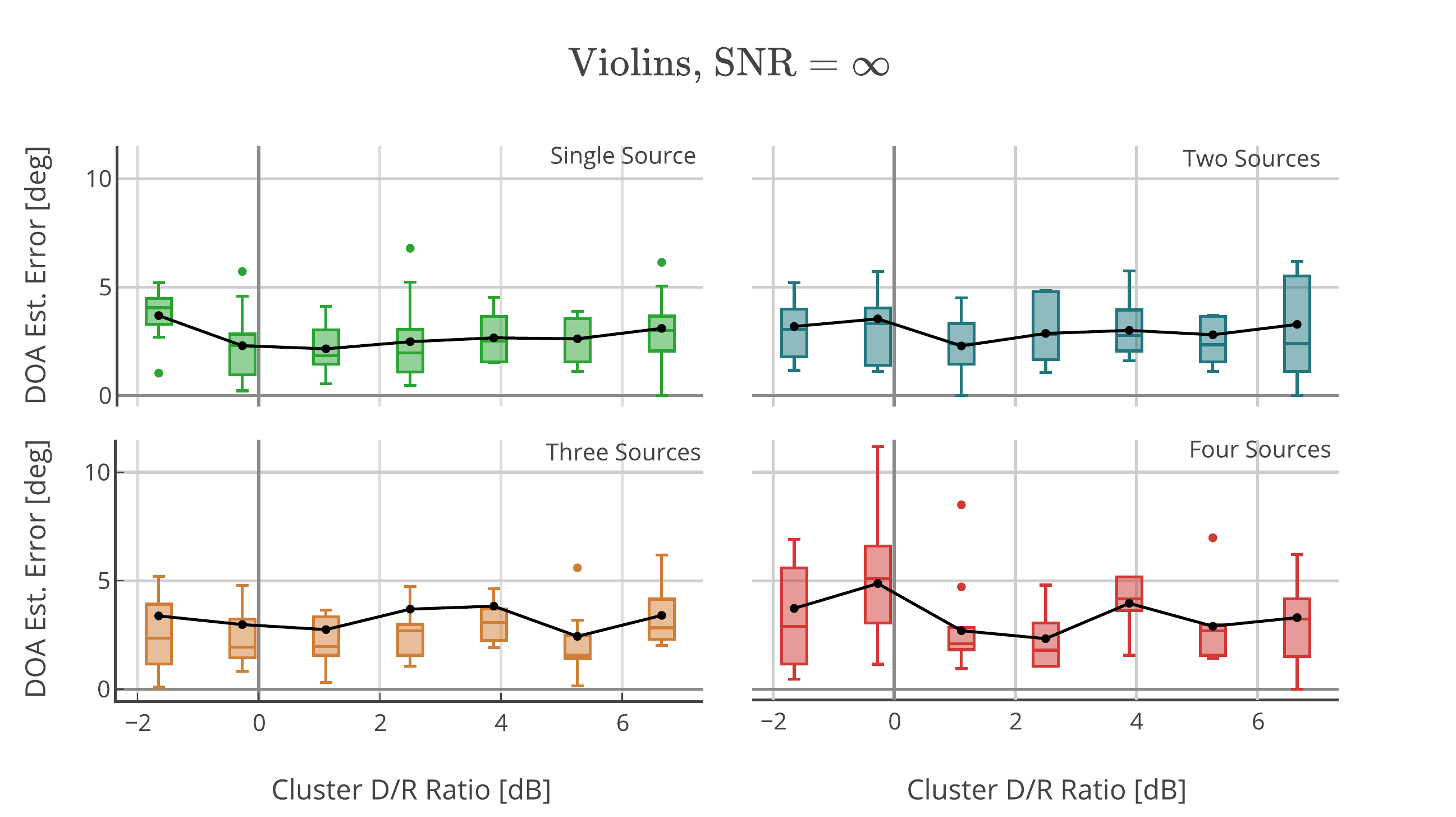}
\caption{Results of DOA estimation for violin signals using HiGRID for $S=1$, $2$, $3$, and $4$ sources and different D/R ratio levels. The box plots indicate the distribution of error. The solid lines indicate the mean DOA error. Asterisks indicate outliers. Extreme values are not shown.}\label{fig:resultsviolin}
\end{figure}

Fig.~\ref{fig:resultsviolin} shows the DOA estimation errors for the case where violin signals are used. There were $5$ extreme values in $280$ distinct test scenarios and $700$ source positions which were not included in the calculation of the average DOA estimation errors. The average DOA estimation errors for $S=1$, $2$, $3$, and $4$ sources were $3.34^{\circ}$, $3.44^{\circ}$, $3.97^{\circ}$, and $4.37^{\circ}$, respectively. The overall average was $3.96^{\circ}$. Similar to the case with speech signals, the DoA estimation error for violin signals is not affected by the D/R ratio or by the number of sources in the scene.

The average number of sources identified for $S=1$, $2$, $3$, and $4$ source cases were, $S_{\textrm{avg}}=1.67$, $2.97$, $3.52$, $4.65$ for speech, and $1.02$, $2.01$, $2.78$, $3.52$ for violin signals, respectively. Negative source disparity was low when D/R ratio was high and also when the source count was low. For speech signals, the the worst performance occured for the low D/R ratio of $-0.27$ dB and for $S=4$. The the average number of sources identified for this condition was $5.7$. For violin signals, the worst performance occured for the moderate D/R ratio of $1.11$ dB and for $S=4$. The average number of sources identified for this condition was $3.2$. Total negative source disparity was observed $8$ times for speech and $47$ times for violins over the total of $280$ test scenarios each.

\subsubsection{Effect of noise} 

\begin{figure}[!t]
\centering
\includegraphics[width=.5\textwidth]{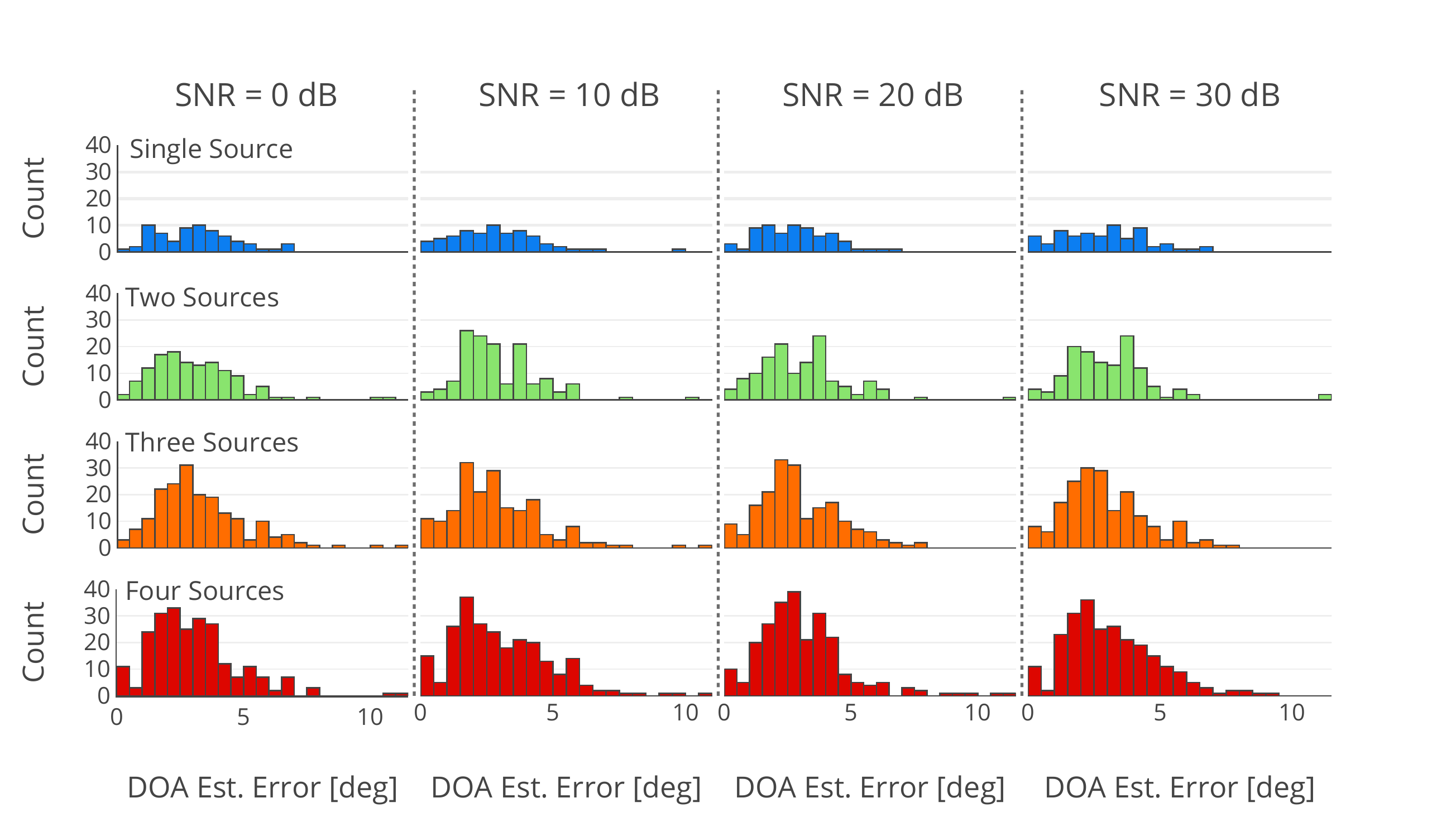}
\caption{Histograms of DOA estimation errors for different number of sources and different noise levels. Bin size is $0.5^{\circ}$ and each count corresponds to the DOA estimation error for a single source. }\label{fig:doasnr}
\end{figure}

Sensor noise was simulated by adding uncorrelated white Gaussian noise to individual microphone signals for different signal-to-noise ratios (SNR). The noise level was calculated based on the energy of the omnidirectional component, obtained as the average of each of the individual channels. The tested levels were $0$, $10$, $20$, $30$ dB SNR. Only the violin signals were used in this evaluation and the selection of the scenarios was made in the same way as in the ideal case.

Fig.~\ref{fig:doasnr} shows the distribution of the error for different number of sources and different noise levels tested. The rows in the figure indicate grouping according to the number of sources and the columns indicate grouping according to the noise level. A one-way analysis of variance (ANOVA) was carried out to understand the effect of noise level on the DOA estimation error. The independent variable was SNR which has four levels from $0$ to $30$ dB. The dependent variable was the DOA estimation error. Results of the ANOVA was significant at $\alpha=0.05$ level ($F(3,2564)=3.625$, $p=0.013$). Levene's test indicated that the variance was different across the tested groups ($F(3,2564)=3.796$, $p=0.01$). Post-hoc comparisons using Games-Howell test indicated that the difference between the error means between $0~\textrm{dB}$ SNR and $10~\textrm{dB}$ SNR as well as $0~\textrm{dB}$ SNR and $20~\textrm{dB}$ SNR are significant. However, these differences (i.e. $0.51^{\circ}$ and $0.49^{\circ}$, respectively) are negligible in a practical sound source localisation scenario. The average number of identified sources and the negative source disparity also did not change with noise. Table~\ref{tab:avgsnr} shows the average number of localised sources. It may be observed that for lower source counts (i.e. $S=1$ and $S=2$) the source count is slightly overestimated in average, and vice versa. Total negative source disparity was $46$, $44$, $45$, $43$ for SNRs of $0$, $10$, $20$, and $30$ dB, over $280$ trials each, respectively. Therefore, the proposed method is considered to be robust to additive, uncorrelated Gaussian white noise.

\begin{table}[!t]
\centering
\caption{Average number of localised sources for different SNRs}
\label{tab:avgsnr}
\begin{tabular}{|
>{\columncolor[HTML]{C0C0C0}}c |c|c|c|c|}
\hline
SNR (dB) & \cellcolor[HTML]{C0C0C0}$S=1$ & \cellcolor[HTML]{C0C0C0}$S=2$ & \cellcolor[HTML]{C0C0C0}$S=3$ & \cellcolor[HTML]{C0C0C0}$S=4$ \\ \hline
$0$        & $1.22$                        & $2.08$                        & $2.92$                        & $3.57$                        \\ \hline
$10$       & $1.01$                        & $2.30$                        & $2.74$                        & $3.75$                        \\ \hline
$20$       & $1.00$                        & $2.12$                        & $2.78$                        & $3.38$                        \\ \hline
$30$       & $1.05$                        & $2.16$                        & $2.84$                        & $3.65$                        \\ \hline
$\infty$ & $1.02$                        & $2.01$                        & $2.78$                        & $3.52$                        \\ \hline
\end{tabular}
\end{table}

\subsection{Comparison of HiGRID with state-of-the-art methods}
We compared HiGRID with three other state-of-the-art methods as well as SRP-based DOA estimation. The state-of-the-art methods chosen for comparison are PIV~\cite{Jarrett:2010vs}, SSPIV~\cite{moore2017direction}, and DPD-MUSIC~\cite{nadiri2014localization}. A general overview of the compared methods as well as the specific test setup and results are presented in this section. Specific details of the algorithms are available in the respective articles cited in the text. We use the SRP method based on plane-wave decomposition as the baseline. The parameters used for PIV, SSPIV and DPD-MUSIC methods (e.g. time and frequency resolutions, amount of overlap in the time-frequency representation, as well as the employed thresholds) in the comparison are selected to match those in the original publications. 

\subsubsection{DOA estimation methods used in the comparison}

PIV method is based on the concept of active intensity which is the vector denoting the net flow of energy in free field~\cite{fahy1995sound}. While it is impossible to measure active intensity using a rigid spherical microphone array, an approximation called pseudo-intensity vector (PIV) can be obtained by using the zeroth and first-order SHD coefficients. Assuming that the sources are W-disjoint orthogonal, their DOAs can be estimated from the long-term directional statistics of PIVs. In the comparisons, the frequency range was between $0$ and $4$ kHz, the window size was $8$ ms, and the overlap was $4$ ms~\cite{Jarrett:2010vs}.




SSPIV extends the PIV method by utilizing higher-order SHD coefficients. SSPIV is calculated at each time-frequency bin by decomposing the time-frequency smoothed spatial covariance matrix obtained from the SHD coefficients into signal and noise subspaces via singular value decomposition (SVD). SSPIV vector is calculated from the first three elements of the signal subspace vector~\cite{moore2017direction}. In the comparisons, the frequency range was between $500$ and $3850$ Hz, the window size was $8$ ms, the overlap was $6$ ms, and the decomposition order was $N=3$~\cite{Jarrett:2010vs}.


DPD-MUSIC is a subspace-based DOA estimation method that uses EB-MUSIC for estimating DOAs in time-frequency bins with a single dominant component~\cite{nadiri2014localization}. While there are recent extensions of DPD-MUSIC that aim to improve estimation accuracy~\cite{7953332} or reduce computational cost~\cite{Rafaely:2017cj} we chose the coherent version of the original algorithm for our comparison since these extensions are originally intended for single-source scenarios. In the comparisons, the frequency range for DPD-MUSIC was between $500$ and $3875$ Hz, the window size was $16$ ms, the overlap was $75\%$, and the decomposition order was $N=3$~\cite{nadiri2014localization}. EB-MUSIC spectrum was calculated at all the pixel centroids of a third-level HEALPix grid and the DOA was estimated for each time-frequency bin as the pixel centroid at which this spectrum is maximum.


For both SRP and HiGRID, the frequency range was between $2608$ and $5214$ Hz, the window size was $21.3$ ms, the overlap was $20$ ms, and the decomposition order was $N=4$. SRP was calculated at all the pixel centroids of a third-level HEALPix grid. For HiGRID, the maximum depth of the analysis tree was $3$. For both methods, DOA estimation was made using the NNL method explained above.



\subsubsection{Test Scenario}
A simple scenario involving four concurrent sources positioned at S1: $(90^{\circ}, 45^{\circ})$, S2: $(90^{\circ}, 135^{\circ})$, S3: $(90^{\circ}, 225^{\circ})$ and, S4: $(90^{\circ},  315^{\circ})$ at a distance of $1.41$ m was emulated using the AIR measurements described above. The D/R ratios for these specific source positions were, $3.04$, $3.29$, $3.56$, and $2.39$ dB, respectively. This constitutes a simple setup with spatially well-separated sources and a high D/R ratio. 

Two types of sources (speech and violin) that were used in the previous section for evaluating the HiGRID approach were also used in the comparison. As with the previous section, the sampling rate and the duration were $F_s=48$ kHz and $4$ s, respectively. In the case involving speech signals sources 1 to 4 are female speech (English), male speech (English), female speech (Danish), and male speech (Danish), respectively. 

\subsubsection{Results}
\begin{table}[!t]
    \centering
    \caption{DOA estimation errors for four concurrent speech sources using different methods.}\label{tab:tab1}
    \begin{tabular}{|c||c|>{\columncolor[HTML]{EFEFEF}}c|c|c|c|}
\hline
\textbf{Source}  & \textbf{SRP}   & \textbf{HiGRID} & \textbf{PIV}    & \textbf{SSPIV} & \textbf{DPD-MUSIC} \\ \hline\hline
\textbf{1}      & $1.54^{\circ}$ & $1.15^{\circ}$  & $13.12^{\circ}$ & $0.87^{\circ}$ & $0.69^{\circ}$     \\ \hline
\textbf{2}      & $1.87^{\circ}$ & $1.34^{\circ}$  & $5.76^{\circ}$  & $1.21^{\circ}$ & $2.04^{\circ}$     \\ \hline
\textbf{3}      & $1.39^{\circ}$ & $0.32^{\circ}$  & $6.86^{\circ}$  & $1.56^{\circ}$ & $1.17^{\circ}$     \\ \hline
\textbf{4}      & $0.80^{\circ}$ & $0.36^{\circ}$  & $3.51^{\circ}$  & $1.28^{\circ}$ & $0.56^{\circ}$     \\ \hline\hline
\textbf{Average} & $1.40^{\circ}$ & \cellcolor[HTML]{9B9B9B} $0.79^{\circ}$  & $7.31^{\circ}$  & $1.23^{\circ}$ & $1.12^{\circ}$     \\ \hline
\end{tabular}
\end{table}

\begin{table}[!t]
    \centering
    \caption{DOA estimation errors for four concurrent violin sources using different methods.}\label{tab:tab2}
    \begin{tabular}{|c||c|>{\columncolor[HTML]{EFEFEF}}c|c|c|c|}
\hline
\textbf{Source}  & \textbf{SRP}   & \textbf{HiGRID} & \textbf{PIV}    & \textbf{SSPIV} & \textbf{DPD-MUSIC} \\ \hline\hline
\textbf{1}      & $1.21^{\circ}$ & $1.15^{\circ}$  & $4.89^{\circ}$  & $6.60^{\circ}$ & $3.45^{\circ}$     \\ \hline
\textbf{2}      & $1.05^{\circ}$ & $1.34^{\circ}$  & $10.07^{\circ}$ & $4.11^{\circ}$ & $4.21^{\circ}$     \\ \hline
\textbf{3}      & $1.15^{\circ}$ & $0.70^{\circ}$  & $20.41^{\circ}$ & $2.31^{\circ}$ & $2.28^{\circ}$     \\ \hline
\textbf{4}      & $1.05^{\circ}$ & $1.18^{\circ}$  & $11.92^{\circ}$ & $7.20^{\circ}$ & $1.12^{\circ}$     \\ \hline\hline
\textbf{Average} & $1.12^{\circ}$ & \cellcolor[HTML]{9B9B9B} $1.09^{\circ}$  & $11.82^{\circ}$ & $5.06^{\circ}$ & $2.77^{\circ}$     \\ \hline
\end{tabular}
\end{table}

DOA estimation results for the two different source types are shown in Tables \ref{tab:tab1} and \ref{tab:tab2}. While calculating the DOA estimation errors, mapping between the identified sources and the ground truth was made using the Kuhn-Munkres algorithm~\cite{kuhn1955hungarian}.

In both scenarios, HiGRID showed a performance that is very similar to SRP with consistent DOA estimates and very small estimation errors both for individual sources and on average. This indicates a distinct advantage over SRP in that the computational cost associated with HiGRID is significantly lower as discussed above. 

It may be observed from Table~\ref{tab:tab1} showing the DOA estimates for speech signals that except for PIV, all methods provide an acceptable level of accuracy with DOA estimation errors in the order of one degree. Overall, PIV method provides the worst  performance while HiGRID provides the best performance. However, the differences between SRP, HiGRID, SSPIV and DPD-MUSIC are not so high as to be significant in a practical context.

The results in Table~\ref{tab:tab2} show the detrimental effects of coherent sources on PIV and to a lesser extent on SSPIV and DPD-MUSIC. Except for HiGRID and SRP, the average DOA estimation errors are much higher than the test using speech signals. HiGRID and SRP provide a DOA estimation performance that is similar to the case with speech sources. However, the performance of PIV and SSPIV are reduced sharply where the DOA estimation from PIV will be completely unusable for Source 3. DOA estimates from DPD-MUSIC for individual sources are all higher than those for the first test case. The average DOA estimation error is larger than that in the first case by a factor of $2.5$.

\subsection{Real Recording}
\begin{figure}[!t]
\centering
\includegraphics[width=.48\textwidth]{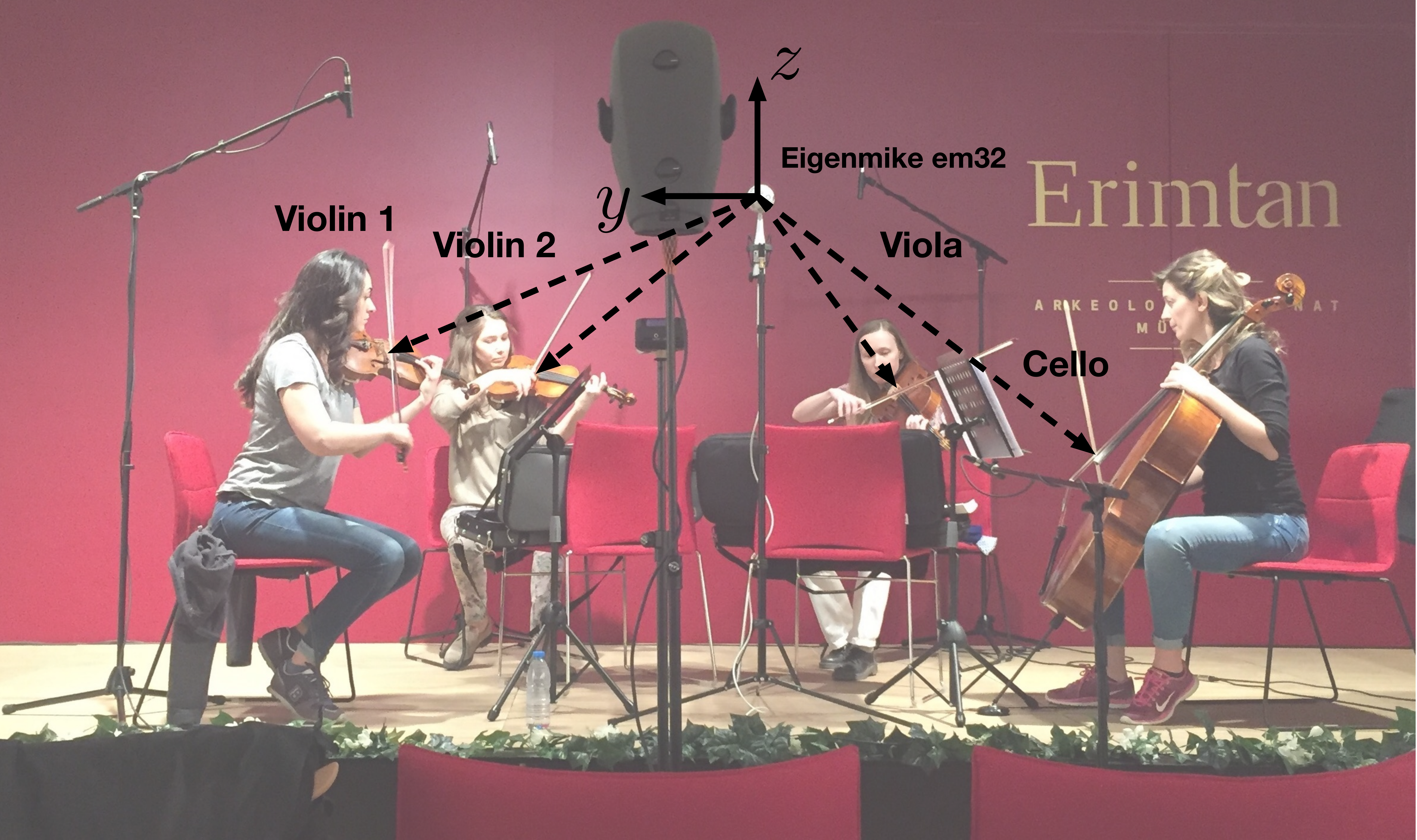}
\caption{Setup for the recording of the classical quartet. Position of the microphone array and directions of the instruments with respect to the array are indicated.}\label{fig:nemeth}
\end{figure}

In order to test the performance of HiGRID in real life conditions, we recorded the pre-concert rehearsal of Nemeth Quartet (a classical quartet consisting of two violins, a viola, a cello) in the small recital hall of Erimtan Museum in Ankara. The reverberation time of the hall was $T_{60}=1.19$ s. The recording was made by positioning the spherical array at a central position between the musicians. The height of the microphone array was chosen as $1.5$ m in order to allow the musicians to maintain eye contact. We were not able to make any precise measurements of the positions of the sound sources due to practical considerations (e.g. movement of musicians during the performance). However, we made AIR measurements with a Genelec 6010A loudspeaker at positions that roughly coincide with the positions of the musical instruments. In order to assess the accuracy of the proposed method, SRP based DOA estimates obtained from windowed direct-path components of AIRs were used as reference. The program material used in the evaluation was a five second excerpt from the third movement of Ludwig van Beethoven's String Quartet Nr.~11, Op.~95. All instruments were playing in this section but the individual parts were not in unison unlike the coherent source scenario simulated in the previous section. Fig.~\ref{fig:nemeth} shows the recording setup.

HiGRID was used with a maximum tree level of $3$. The maximum order of SHD was $N=4$. Two different frequency ranges were used: 1) $2608$ Hz to $5216$ Hz, and 2) $1308$ Hz to $5216$. A $1024$ point FFT with a hop size of $64$ samples was used for the STFT. The reference DOAs and estimation results are given in Table~\ref{tab:nemeth}. It may be observed that using the first frequency range, three out of four instruments (two violins and the viola) can be localised. No spurious sources were detected in this case. Using the second frequency range, all instruments can be localised along with six spurious DOAs corresponding to strong room reflections (not shown in the table). It should be noted that reporting the DOA estimation error with respect to the reference values would be misleading due to the inaccuracy of the reference DOAs as well as the sources being volumetric and not static. Therefore, the presented results are only intended to show --with reasonable but imprecisely quantified accuracy-- that HiGRID can find source DOAs under real-life conditions.

\begin{table*}[]
\centering
\caption{Reference and estimated DOAs for the microphone array recording of the classical quartet}
\label{tab:nemeth}
\begin{tabular}{|c|c|c|c|c|}
\hline
\rowcolor[HTML]{C0C0C0} 
\cellcolor[HTML]{C0C0C0}$(\theta,\phi)$  & \textbf{Violin 1}                        & \textbf{Violin 2}                        & \textbf{Viola}                            & \textbf{Cello}                            \\ \hline
\cellcolor[HTML]{C0C0C0}\textbf{Reference}        & $(116.1^{\circ}, 92.4^{\circ})$ & $(114.9^{\circ}, 32.3^{\circ})$ & $(124.4^{\circ}, 327.8^{\circ})$ & $(132.4^{\circ}, 268.4^{\circ})$ \\ \hline
\cellcolor[HTML]{C0C0C0}$2608 - 5216$ Hz & $(109.7^{\circ}, 89.4^{\circ})$ & $(109.7^{\circ}, 28.1^{\circ})$ & $(112.3^{\circ}, 340.4^{\circ})$  & -                                \\ \hline
\cellcolor[HTML]{C0C0C0}$1304 - 5216$ Hz & $(118.1^{\circ}, 88.5^{\circ})$ & $(107.1^{\circ}, 30.9^{\circ})$ & $(115.0^{\circ}, 337.7^{\circ})$ & $(138.1^{\circ}, 276.4^{\circ})$ \\ \hline
\end{tabular}
\end{table*}

\section{CONCLUSIONS}\label{sec:conc}
Estimation of DOAs of multiple, possibly coherent sources in highly reverberant environments is an important step in sound scene analysis. While there exist different methods proposed for this purpose, these are either computationally prohibitively costly, or they fail for highly coherent sound sources. An extension to steered response power (SRP) and an accompanying entropy-based hierarchical grid refinement method (HiGRID) was proposed in this article. The extension of SRP, which we call steered response power density (SRPD) uses the spherical harmonic decomposition of a sound field to extract the power density in a given region on the unit sphere. This allows representing the probability of a source being present in the given analysis region on the unit sphere. SRPD is calculated on a HEALPix grid which is a spherical tessellation allowing a multiresolution representation of data. Maxima of the SRPD map is searched on this grid using information gain as a decision measure. The obtained multiresolution maps are segmented to obtain multiple local DOA estimates. These are then aggregated to estimate DOAs of multiple sources. We showed that HiGRID has a lower computational cost than SRP, is robust to reverberation and additive noise, and can accurately localise highly coherent sources as well as speech. HiGRID also compared favourably with other state of the art methods in terms of its DOA estimation accuracy. Application of HiGRID using a microphone array recording of a musical performance demonstrated its utility under real-life conditions.


\section*{Acknowledgment}
We would like to thank the members of Nemeth Quartet (\url{http://www.nemethquartet.com}) for allowing us to record their rehearsal. This paper is dedicated to the memory of the third author's late grandmother, Belk\i{}s Hac\i{}habibo\u{g}lu.

\ifCLASSOPTIONcaptionsoff
  \newpage
\fi



%
\bibliographystyle{IEEEtran}
\bibliography{refs17}
%
\newpage
\begin{IEEEbiographynophoto}
{Mert Burkay \c{C}\"{o}teli} (S'18) received his B.Sc. (honors) and M.S. in electrical and electronic engineering from the Middle East Technical University (METU), Ankara, Turkey, in 2009 and 2013, respectively. He is currently pursuing his Ph.D. in Information Systems at the Graduate School of Informatics at the same university. He is also working as a senior system engineer at Aselsan A.S. His research interests include microphone array signal processing and acoustic scene analysis.   	
\end{IEEEbiographynophoto}



\begin{IEEEbiographynophoto}
{Orhun Olgun} (S'18) was born is Ankara, Turkey in 1994. He received the B.Sc degree in Electrical and Electronics engineering from Bilkent University, Ankara, Turkey in 2016. He is currently pursuing his M.Sc. in Multimedia Informatics at the Graduate School of Informatics, METU, Ankara, Turkey. His research interests include 3D audio, microphone array signal processing, and acoustic scene analysis.
\end{IEEEbiographynophoto}

\begin{IEEEbiographynophoto}
{H\"{u}seyin Hac\i{}habibo\u{g}lu} (S'96-M'00-SM'12) is an Associate Professor of Signal Processing and Head of Department of Modelling and Simulation at Graduate School of Informatics, Middle East Technical University, Ankara, Turkey. He received the B.Sc. (honors) degree from the Middle East Technical University (METU), Ankara, Turkey, in 2000, the M.Sc. degree from the University of Bristol, Bristol, U.K., in 2001, both in electrical and electronic engineering, and the Ph.D. degree in computer science from Queen's University Belfast, Belfast, U.K., in 2004. He held research positions at University of Surrey, Guildford, U.K.\ (2004-2008) and King's College London, London, U.K.\ (2008-2011). His research interests include audio signal processing, room acoustics, multichannel audio systems, psychoacoustics of spatial hearing, microphone arrays, and game audio. Dr.\ Hac\i{}habibo\u{g}lu is a Senior Member of the IEEE, Mmmber of the IEEE Signal Processing Society, Audio Engineering Society (AES),  Turkish Acoustics Society (TAD), and the European Acoustics Association (EAA) and the associate editor of \textit{IEEE/ACM Transactions on Audio, Speech, and Language Processing}.	
\end{IEEEbiographynophoto}
\vfill







\end{document}